%% file: draft_second.tex
\documentclass[12pt,preprint,letterpaper,nofootinbib,superscriptaddress, preprintnumbers]{revtex4-1}

\usepackage[font=small,labelfont=bf]{caption}
\usepackage{subcaption}
\usepackage{graphicx} 
\usepackage{epstopdf}
\usepackage{bm}
\usepackage[colorlinks=true,linktocpage=true,linkcolor=blue,citecolor=blue,urlcolor=blue]{hyperref}
\usepackage{textcomp}
\usepackage{color}
\usepackage{xcolor}
\usepackage{tabularx}
\usepackage{tabu}
\usepackage{amsmath,amssymb,mathtools}
\usepackage[normalem]{ulem}

\usepackage{bbold}
\usepackage{tikz}
\usepackage{tikz-feynman}
\usepackage{contour}
\usepackage{adjustbox}
\usepackage{orcidlink}
\usepackage{cancel}

\definecolor{lime}{HTML}{A6CE39}

\def \MET{\rm E{\!\!\!/}_T}
\newcommand{\opdhphi}{\mathcal{O}_{DH\partial\phi}}
\newcommand{\opdhchi}{\mathcal{O}_{DH\chi\chi}}
\newcommand{\opephi}{\mathcal{O}_{e\phi}}
\newcommand{\opuphi}{\mathcal{O}_{u\phi}}
\newcommand{\opdphi}{\mathcal{O}_{d\phi}}

\newcommand{\wcdhphi}{C_{DH\partial\phi}}
\newcommand{\wcdhchi}{C_{DH\chi\chi}}
\newcommand{\wcephi}{C_{e\phi}}
\newcommand{\wcuphi}{C_{u\phi}}
\newcommand{\wcdphi}{C_{d\phi}}

\newcommand{\wc}{C}
\newcommand{\op}[1]{\mathcal{O}_{#1}}

\begin{document}

\title{ Semi-visible higgs decay as a probe for new invisible particles}

\def\BNL{High Energy Theory Group, Physics Department, Brookhaven National Laboratory, Upton, NY 11973, USA}
\def\MON{
	School of Physics and Astronomy, Monash University, Wellington Road, Clayton, Victoria 3800, Australia}

\author{Sally Dawson\orcidlink{0000-0002-5598-695x
}}
\email{dawson@bnl.gov}
\affiliation{\BNL}
\author{Arnab Roy\orcidlink{0000-0003-0249-1440
}}
\email{arnab.roy1@monash.edu}
\affiliation{\MON}
\author{German Valencia\orcidlink{0000-0001-6600-1290
}}
\email{german.valencia@monash.edu}
\affiliation{\MON}

\begin{abstract}
\noindent
    We discuss the HL-LHC sensitivity to probe new invisible particles including scalars and fermions using semi-visible Higgs decays in the $pp\to ZH, Z\to jj\, (\ell^+\ell^-), ~H\to \ell^+\ell^- (jj) + \MET$ production mode. The kinematics of these decays allow new particle masses below $m\lesssim 50$~GeV. We carry out our analysis using both a cut-based approach and a multivariate method based on a boosted decision tree. We work within the dark-SMEFT framework with operators up to dimension six and a discrete $\mathbb{Z}_2$ symmetry under which the new particles are odd and the SM particles are even. We compare our results to those obtained from considering the invisible $Z$-width, as well as perturbative unitarity arguments. Finally, we outline kinematic strategies at the LHC to distinguish different operator structures of the postulated invisible particles. 
\end{abstract}

\maketitle

\section{Introduction}

Semi-visible Higgs decays into a pair of leptons or two jets plus missing transverse energy ($\MET$) provide a window into the production of new invisible particles that complements other channels~\cite{Aguilar-Saavedra:2022xrb}. We discuss these processes assuming that the $\MET$ is due to a pair of new light particles with  $m\lesssim 50$~GeV, which can have spin zero or spin one-half.\footnote{There is minimal sensitivity for the mass range between 50~GeV and the kinematic limit near $m_H/2$.} We carry out the analysis using a dark-SMEFT framework including operators up to dimension six and with a discrete $\mathbb{Z}_2$ symmetry under which the new particles are odd and the Standard Model (SM) particles are even. The fields corresponding to the new states are taken to be complex, thus allowing for the possibility of them carrying a dark charge.
Specifically, we concentrate on $ZH$ production at the LHC followed by the semi-visible decay $H\to \ell^+\ell^-+\MET$ or $H\to jj+\MET$. Even though the gluon fusion channel for Higgs production has a larger cross-section,  we choose to study the $ZH$ channel because it leads to cleaner signals \cite{ATLAS:2018nda,Aguilar-Saavedra:2022xrb}. 

We begin by listing the operators that we will include in our study in Section \ref{sec:framework} and derive the theoretical constraints on their Wilson coefficients that follow from perturbative unitarity in Section \ref{sec:unitarity}. In Section \ref{sec:z-width} we consider the invisible $Z$ width, which can constrain some of our Wilson coefficients. Our new analysis for semi-visible Higgs decay is described in  Section \ref{sec:analysis}, and we present our results in Section \ref{sec:results}, where we discuss the fate of the Higgs-neutrino floor~\cite{Aguilar-Saavedra:2022xrb} in this context, and contrast the sensitivity reach of the semi-visible Higgs processes with those obtained from the invisible $Z$ width and from  perturbative unitarity considerations.

\section{EFT framework}
\label{sec:framework}

We consider scenarios where a pair of Beyond the Standard Model (BSM) invisible particles interact with SM particles through $SU(3)\times SU(2)\times U(1)$ gauge-invariant dimension-6 operators. The possible set of  operators, dubbed DSMEFT operators, have been catalogued before in the literature ~\cite{Criado:2021trs,Aebischer:2022wnl} with several subsets considered for specific cases \cite{Harnik:2008uu,Goodman:2010yf,Goodman:2010qn,Beltran:2010ww,Goodman:2010ku,Kamenik:2011vy,Fox:2011pm,DelNobile:2011uf,Cheung:2012gi,Buckley:2013jwa,Bell:2013wua,DeSimone:2013gj,Fedderke:2014wda,Matsumoto:2014rxa,Crivellin:2014qxa,Crivellin:2014gpa,Duch:2014xda,Hisano:2015bma,Crivellin:2015wva,Matsumoto:2016hbs,DeSimone:2016fbz,Bishara:2016hek,Bauer:2016pug,Bruggisser:2016nzw,Pobbe:2017wrj,Belwal:2017nkw,Han:2017qkr,Brod:2017bsw,Belyaev:2018pqr,Arina:2020mxo,Barman:2020plp,Barman:2020ifq,Bhattacharya:2021edh,GAMBIT:2021rlp,Barducci:2021egn,He:2022ljo,Borah:2024twm,Roy:2024ear,Roy:2025pht,Borah:2025ema}.
The operators relevant for our analysis follow from requiring a $\mathbb{Z}_2$ symmetry under which all BSM fields are odd and all SM fields are even, and by selecting only those operators that can contribute to semi-visible Higgs decays. The new particles containing such a stabilizing $\mathbb{Z}_2$ symmetry can be complex scalars ($\phi$) or Dirac fermions ($\chi$) and are often motivated as possible dark matter (DM) candidates. For this reason, we use `DM' as a general acronym for the new invisible particles.

We note that operators involving multiple SM  Higgs bosons and DM fields without any other SM fields or covariant derivatives are strongly constrained by the measurements of Higgs invisible decays~\cite{CMS:2022qva,ATLAS:2022yvh} or from direct detection experiments~\cite{Arcadi:2021mag}. The operator $\mathcal{O}_{\Box\phi}^{(6)}=(H^\dagger H) \Box\, \phi^\dagger\phi$ is an exception~\cite{Ruhdorfer:2024dgz} to the direct detection limits, since this operator yields momentum-suppressed DM–nucleon interactions; however, the Higgs invisible-decay constraints on $\mathcal{O}_{\Box\phi}^{(6)}$ are applicable, and they are much stronger than those from semi-visible modes. \footnote{Applying the constraint placed by the invisible Higgs width on $C_{\Box\phi}$,  $|C_{\Box\phi}|\lesssim 0.55~\text{TeV}^{-2}$ \cite{Criado:2021trs}
    results in ${\cal B}(h\to \mu^+\mu^-\MET)\lesssim 10^{-17}$, well below what can be explored with semi-visible Higgs decay.} Therefore, we exclude these purely Higgs-portal operators from our study. 

The relevant set of operators for scalar DM fields is then,
\begin{align}
\op{DH\partial\phi} &= (H^\dagger \overleftrightarrow{D}^\mu H)(\phi^\dagger \overleftrightarrow{\partial_\mu}\phi)  
& \op{e\phi}{}  &= (\overline l_i e_j H) \phi^\dagger\phi \\
\op{u\phi}{} &= (\overline q_i u_j \widetilde H) \phi^\dagger\phi &  \op{d\phi}{} &= (\overline q_i\, d_j H) \phi^\dagger\phi,
\label{eq:Soperators}
\end{align}
where the double arrow notation has the usual meaning, e.g. $(\phi^\dagger \overleftrightarrow{\partial_\mu}\phi)\equiv (\phi^\dagger \overrightarrow{\partial_\mu}\phi)-(\phi^\dagger \overleftarrow{\partial_\mu}\phi)$.
The fermion DM operators are, 
\begin{align}
\op{DH\chi\chi}{} &=  (iH^\dagger \overleftrightarrow{D}^\mu H)  (\overline \chi\gamma_\mu  \chi) &
\op{DH\chi\chi 2}{} &=  (iH^\dagger \overleftrightarrow{D}^\mu H)  (\overline \chi\gamma_\mu\gamma^5  \chi),
\label{eq:Xoperators}
\end{align}
where $H$ is the SM Higgs doublet, $l_i, q_i$ are the left-handed lepton and quark doublets, with flavor index $i$, and $e_i,u_i,d_i$ are the right-handed charged lepton, up- and down-quarks with flavor index $i$, respectively. We assume that all DSMEFT operators are diagonal in flavor space and independent of the flavor. 

The dimension-6 DSMEFT Lagrangian that we consider is,
\begin{align}
L_{DSMEFT}=L_{SM}+\sum_\alpha{C_\alpha\over \Lambda^2}{\cal{O}_\alpha},
\end{align}
where $L_{SM}$ is the SM Lagrangian, the dimension-6 DSMEFT operators are defined in Eqs. \eqref{eq:Soperators} and \eqref{eq:Xoperators}, and $\Lambda$ is the scale of new physics.  All information about the new physics is contained in the coefficient functions $C_\alpha$ , considered to be real without any loss of generality. 

For operators involving new particles, there is no interference between the SM and the EFT operators, which results in observables being proportional to $|C_\alpha/\Lambda^2|^2$. Therefore, the imaginary parts of these coefficients receive the same constraints as the corresponding real parts. Pure dimension-6 SMEFT operators constructed solely from SM fields are not included in the present analysis, because their contributions to Higgs semi-visible decay are further suppressed by additional powers of $\Lambda$. As the BSM interaction enters only in the Higgs decay, the relevant EFT expansion parameter is set by Higgs-decay kinematics, $E\sim M_H$, leading to higher order terms being negligible for $\Lambda$ in the multi-TeV range.

We do not consider spin-1 dark states ($X$) at dimension six. When written with vector fields, the nominal dimension-6 couplings produce amplitudes yielding rates that become ill-behaved as $M_X\to 0$. In gauge-invariant completions where $X^\mu$ is a dark gauge boson acquiring mass by some Higgs mechanism \cite{Kamenik:2011vy,Williams:2011qb}, the effective operators inherit a coefficient that vanishes in the limit of massless $X_\mu$. With masses $M_X\lesssim 50$~GeV, as required kinematically for semi-visible Higgs decays in the ZH process, the operators are effectively of higher dimension with coefficients $(M_X/\Lambda)$ for each $X_\mu$ that cannot be rewritten as a field strength tensor. Using field strengths $X^{\mu\nu}$ to define a set of operators, on the other hand, pushes the minimal interactions to dimension- eight given the assumed $\mathbb{Z}_2$ symmetry, leading to suppressed rates at the LHC. For these reasons, vector states are not included in this dimension-6 analysis.

The operators in Eqs.~\eqref{eq:Soperators} and \eqref{eq:Xoperators}, without explicit quark fields lead to semi-visible Higgs decays with leptons in the final state, whereas those with explicit quark fields result in jets in the final state. 

 \section{Perturbative unitarity bounds}
\label{sec:unitarity}

 To ascertain the significance of our results, it is useful to compare them with bounds obtained by requiring that perturbative unitarity for certain processes be satisfied. To do so we first recall that the new physics scale of our effective theory is assumed to be larger than a few TeV. Furthermore, since we use the effective theory to explore $H$ and $Z$ decays, we only need it to be valid up to scales of around a few hundred GeV.

\subsection{Operators with only one Higgs field} 
        For $\op{e\phi}=(\overline l_i e_j H) \phi^\dagger\phi$, we consider tree-level scattering $e^+e^-\to \bar\phi\phi$. The leading partial wave is $\ell =0$, and requiring it to be less than $\frac{1}{2}$ results in the bound \footnote{We employ unitarity bounds including threshold kinematic factors instead of the more common asymptotic bounds as required for our comparisons. See for example~\cite{Endo:2014mja}.},
     \begin{align}
     \frac{\wcephi}{\Lambda^2}<\frac{8\pi}{v\sqrt{M_{\phi\phi}^2}}\left(\frac{M_{\phi\phi}^2-4M_{\phi}^2}{M_{\phi\phi}^2}\right)^{-1/4} ,
     \label{boundoephi}
     \end{align}

where $v$ denotes the usual Higgs vacuum expectation value ($v\simeq246~\mathrm{GeV}$). The corresponding operators with quarks, $\op{u\phi},~\op{d\phi}$ have unitarity bounds that are smaller by a factor of $\sqrt{3}$ relative to that of Eq.~\ref{boundoephi}, which follow from considering the color singlet channel for the two quarks.

\subsection[
Operators with the factor Hdagger Dmu H
]{Operators with the factor $(H^\dagger \protect\overleftrightarrow{D}^{\mu} H)$}

In this case, we consider both tree-level scattering $HZ\to \bar \phi\phi$ and $Z$ mediated $\bar \phi\phi\to \bar \phi\phi$ scattering. For the former, the leading partial wave is $\ell=1$, whereas for the latter it is $\ell=0$. 

For scalar operators, $\op{DH\partial\phi}$, these bounds are respectively
\begin{align}
\label{eq:unitscalar}
\frac{|C_{DH\partial\phi}|}{\Lambda^2} 
&<\frac{24\pi\,M_Z}{g_Z v\,(M^2_{\phi\phi}+M_Z^2-M_H^2)}
		\frac{\lambda^{1/2}(M^2_{\phi\phi},M_Z^2,M_H^2)}{M^2_{\phi\phi}\,\beta_\phi},~ g_Z=\frac{2M_W}{v} \nonumber\\
 \frac{|C_{DH\partial\phi}|}{\Lambda^2} 
&\leq \frac{1}{g_Z v^2}\frac{\sqrt{8\pi}}
{ \left[\left( \frac{(1 + \beta_\phi^2)}{\beta_\phi^2} + \frac{M_Z^2}{M^2_{\phi\phi}\beta_\phi^2} \right) \ln \left( 1 + \frac{M^2_{\phi\phi} \beta_\phi^2}{M_Z^2} \right) - 1 \right]^{1/2}}
\end{align}
where  $\beta_\phi = \sqrt{1 - \frac{4 M_{\phi}^2}{M^2_{\phi\phi}}}$ and $\lambda(a,b,c)=a^2+b^2+c^2-2ab-2ac-2bc$. For fermionic operators, $\op{DH\chi\chi}{}$ and $\op{DH\chi\chi2}{}$,  the unitarity bounds are
\begin{align}
\label{eq:unitfermion}
   \frac{|C_{DH\chi\chi}|}{\Lambda^2} &< \frac{24\pi\,M_Z}{g_Z v\,(M^2_{\chi\chi}+M_Z^2-M_H^2)}
		\frac{\lambda^{1/2}(M^2_{\chi\chi},M_Z^2,M_H^2)}{M^2_{\chi\chi}\beta_\chi}\nonumber \\
     \frac{|C_{DH\chi\chi}|}{\Lambda^2}
&<
\frac{1}{g_Z v^2}
\frac{\sqrt{8\pi}}{
\left[\frac{M_{\chi\chi}^2(1+\beta_\chi^2)}{2(M_{\chi\chi}^2-m_Z^2)}
+1
-\frac{m_Z^2}{M_{\chi\chi}^2\beta_\chi^2}
\ln\!\left(1+\frac{M_{\chi\chi}^2\beta_\chi^2}{m_Z^2}\right)
\right]^{1/2}},
\end{align}
and similarly for $C_{DH\chi\chi 2}$, with $\beta_\chi = \sqrt{1 - \frac{4 M_{\chi}^2}{M^2_{\chi\chi}}}$. These two bounds are similar in magnitude for $M_{\phi\phi} \,(\text{or}~ M_{\chi\chi})\sim 1$ TeV, with the latter being stronger for the smaller values of $M_{\phi\phi}$ due to the threshold factor in the former.

\section{Z-invisible decay width}
\label{sec:z-width}

The operators $\opdhphi$, $\opdhchi$ and ${\cal O}_{DH\chi\chi 2}$ contribute to the invisible decay of the Z boson. We calculate the Z-invisible decay width from these operators with different DM masses using \texttt{Madgraph5} \cite{Alwall:2014hca}. To obtain the corresponding constraints, we require these contributions to the invisible $Z$ width, $\Gamma_{\rm inv}^{\rm new}$,  to be below the difference between the experimental limit and  the SM contribution  ($Z\to \nu{\overline{\nu}}$) at the 95\% confidence level \cite{ALEPH:2005ab},
\begin{align}
  \Gamma_{\rm inv}^{\rm new} \;<\; 2.04 ~\text{MeV} ~~\text{(95$\%$ CL)} .
  \label{eq:z_inv_width}
\end{align}

As a sanity check, we have verified that these contributions to the total Higgs width do not result in Higgs decay branching fractions larger than 20\%, or about twice the current 95\% C.L. upper bound on $BR(H\to {\rm invisible})$ \cite{ATLAS:2023tkt}.

\section{Collider phenomenology}
\label{sec:analysis}

The largest Higgs production cross-section at the LHC originates from gluon fusion, but this process suffers from a large QCD background. For semi-visible modes generating $\MET$, it is therefore better to choose the cleanest production mechanism at the expense of a lower cross-section. This leads us to focus on the conventional channel for this type of study, $pp\to ZH, Z\to jj, ~H\to \ell^+\ell^-\MET$ \cite{Aguilar-Saavedra:2022xrb}, and $pp\to ZH,Z\to\ell^+\ell^-,~H\to jj\,\MET$ for the hadronic DSMEFT operators \cite{Englert:2012wf}, shown in Fig.~\ref{fig:zh_cascades} for illustration. We treat the leptonic- and hadronic semi-visible Higgs decays as two mutually exclusive analyses, despite the common object-level signature $2\ell+2j+E_T^{\rm miss}$. The channels are disentangled by the $Z$-tag assignment: we require the tagging system (either $\ell^+\ell^-$ or $jj$) to reconstruct to the $Z$ mass. The relevant Feynman diagrams for semi-visible Higgs decay mechanisms in the SM and in DSMEFT are shown in Fig.~\ref{fig:feynmandiagrams}.

\begin{figure}[t]
	\fbox{
		\begin{minipage}{0.48\linewidth}
			\centering
			\begin{tikzpicture}[node distance=8mm,>=Latex]
				\node (pp) {$pp$};
				\node[right=10mm of pp] (Z) {$Z$};
				\node[right=13mm of pp] (H) {$H$};
				\draw[->] (pp) -- (Z);
				
				\node[below left=4mm and 8mm of Z] (jj) {$jj$};
				\node[below right=4mm and 8mm of H] (llmet) {$\ell^+\ell^- + \MET$};
				
				\draw[->] (Z) |- node[pos=0.35,left] {} (jj);
				\draw[->] (H) -| node[pos=0.35,right] {} (llmet);
			\end{tikzpicture}
            
			\vspace{3mm}
			\textbf{(1)}\; $pp\to ZH,\; Z\to jj,\; H\to \ell^+\ell^-\,\MET$
		\end{minipage}\hspace{3mm}%
			\begin{minipage}{0.48\linewidth}
			\centering
			\begin{tikzpicture}[node distance=8mm,>=Latex]
				\node (pp) {$pp$};
				\node[right=10mm of pp] (Z) {$Z$};
				\node[right=13mm of pp] (H) {$H$};
				\draw[->] (pp) -- (Z);
				
				\node[below left=4mm and 8mm of Z] (jj) {{$\ell^+\ell^-$}};
				\node[below right=4mm and 8mm of H] (llmet) {$jj + \MET$};
				
				\draw[->] (Z) |- node[pos=0.35,left] {} (jj);
				\draw[->] (H) -| node[pos=0.35,right] {} (llmet);
			\end{tikzpicture}
            
			\vspace{3mm}
			\textbf{(2)}\; $pp\to ZH,\; Z\to \ell^+\ell^-,\; H\to jj\,\MET$
		\end{minipage}
	}
	\caption{Signal final states explored in this study for the semi-visible leptonic (left) and hadronic (right) Higgs decays.}
	\label{fig:zh_cascades}
\end{figure}

\input{feynman}

Using UFO model files generated by \texttt{FeynRules}~\cite{Alloul:2013bka} as input, we generate signal (with the presence of the DSMEFT operators) and background events for $pp\rightarrow ZH, H\rightarrow {\textrm{semi-visible~ final~ states}}$ in \nolinkurl{Madgraph5}~\cite{Alwall:2014hca} at leading order, with the \texttt{PDF4LHC15\_nlo\_mc} parton distribution function set~\cite{Butterworth:2015oua} and $\sqrt{s}=14$ TeV. The renormalization and factorization scales are set to the partonic center-of-mass energy $\sqrt{\hat{s}}$. For the inclusive $t\bar{t}$ background, we employ the MLM approach~\cite{Mangano:2006rw} with $\rm \texttt{XQcut}=20~GeV$ and $\rm \texttt{Qcut}=30~GeV$ to match the partonic jets to those coming from the parton showers~\cite{Alwall:2007fs}.\footnote{
Here \texttt{XQcut} is the minimum matrix-element-level jet measure ($k_T$), while \texttt{Qcut} is the corresponding Pythia shower-level matching scale; together they define the transition between matrix-element partons and parton-shower emissions.
} The parton-level events are then processed with \nolinkurl{Pythia8}~\cite{Sjostrand:2006za,Sjostrand:2007gs} to simulate parton showering and hadronization. Detector effects are incorporated using \nolinkurl{Delphes3}~\cite{deFavereau:2013fsa}, configured with the CMS detector card.

Both categories of the signal final states presented in Fig.~\ref{fig:zh_cascades} contain two leptons, two jets and $\MET$. We select events with $\rm p_T(\ell_1) > 20~GeV$ for the leading lepton and $\rm p_T(\ell_2) > 10~GeV$ for the subleading lepton. In addition, we require the pseudo-rapidity of all leptons in the range $|\eta| < 2.5$. To ensure lepton isolation, we use the energy-flow objects from \nolinkurl{Delphes}, following the prescription of Ref.~\cite{Khachatryan:2016uwr}. The isolation criterion is given by,
\begin{equation}
    \frac{\sum p_{T}^{R<r}}{p_{T,\ell}} < I, \quad \ell = e, \mu,
\end{equation}
where the isolation cone size is fixed to $r = 0.3$, and the threshold $I$ is set to $0.12$ for electrons and $0.25$ for muons. 

We use \texttt{FastJet~3.3.2} \cite{Cacciari:2011ma} to reconstruct jets from the Delphes energy-flow objects, which emulates a particle-flow reconstruction algorithm~\cite{CMS:2009nxa} with the anti-$k_T$ algorithm~\cite{Cacciari:2008gp} and a radius parameter $R = 0.4$. Only jets with $p_T^j > 20~\mathrm{GeV}$ and $|\eta| < 4.0$ are considered. Although b-jets are absent in the signal, they arise in the dominant top-quark background. To identify $b$- jets, we match the reconstructed jets to $b$ quarks at the parton level, requiring them to lie within $|\eta| < 2.5$ and within a cone of $\Delta R < 0.3$ around the jet axis. The b-tagging efficiency ($\epsilon_b^{\rm eff}$) in our study arises solely from this matching procedure, and is essentially the same for all the background processes having top-quarks before applying analysis cuts. After the kinematic selection cuts, $\epsilon_b^{\rm eff}$ varies mildly from process to process (typically $\sim 0.78-0.86$), reflecting selection-induced differences in jet kinematics and event topology. The top-quark backgrounds are reduced by vetoing events containing any reconstructed b-jet.

\subsection{Higgs-neutrino floor}

For the semi-visible Higgs decay $H\to Z+\MET \to 2\ell +\MET$, there is an irreducible background  $H\to ZZ^* \to \ell^+\ell^- \nu{\overline{\nu}}$ dubbed the ``Higgs-neutrino floor", which has been studied for several models that produce $Z+\MET$  \cite{Aguilar-Saavedra:2022xrb,ATLAS:2022yvh,ATLAS:2023tkt,ATLAS:2024itc}.  In our study, we expect this background to affect results for all of the operators with factors of $(H^\dagger \overleftrightarrow{D}^\mu H)$, as these operators result in diagrams such as those in the center panel of Fig.~\ref{fig:feynmandiagrams}. As demonstrated below, however, this background is not dominant in our case. For the remaining operators, with two explicit fermion fields,  the two leptons (or jets) in the final state $2\ell +\MET$ do not reconstruct to the Z boson mass, as shown in the right panel of Fig.~\ref{fig:feynmandiagrams}. All in all, the Higgs-neutrino-floor background is a reducible background in our study. 

\begin{figure}[t]
    \centering
    \begin{subfigure}{0.32\textwidth}
        \centering
        \includegraphics[width=\linewidth]{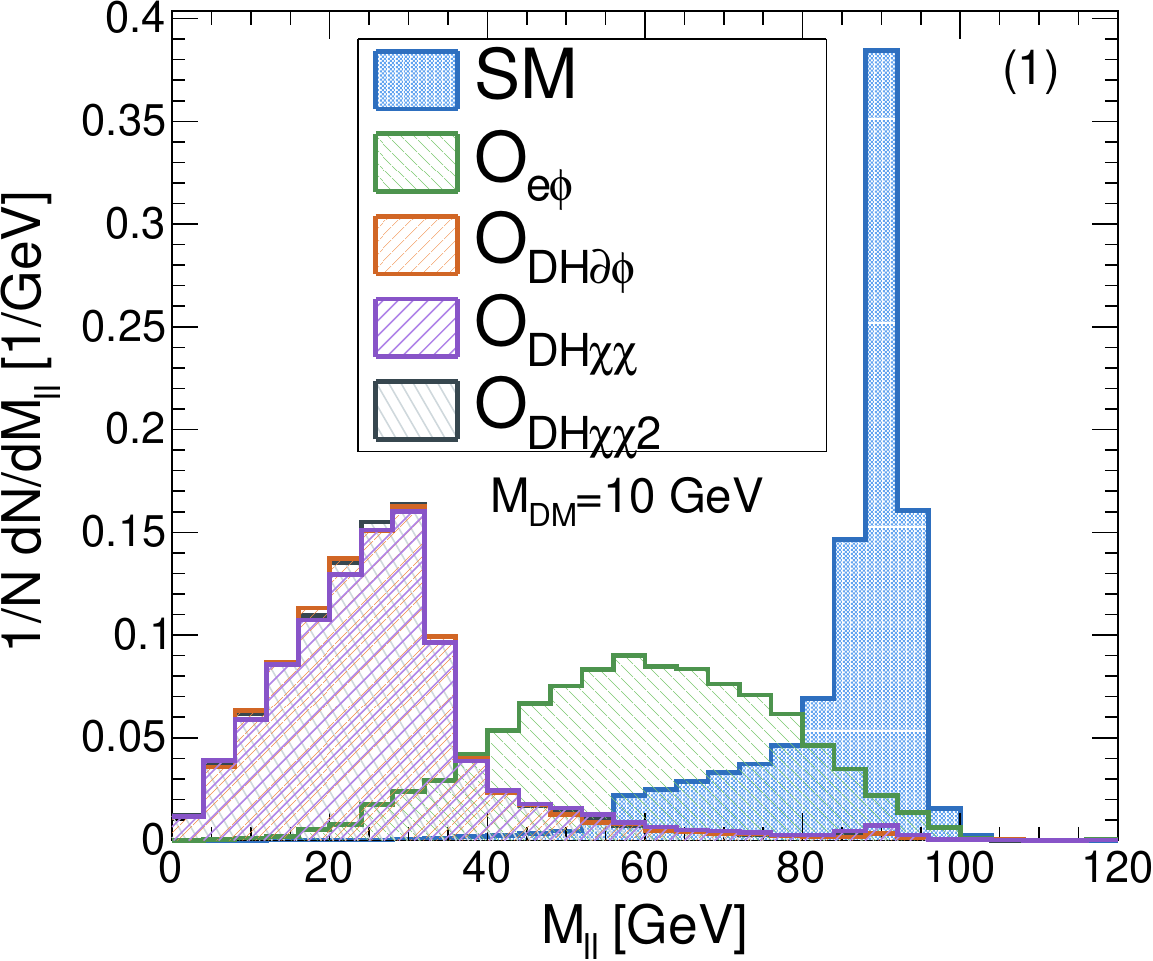}
        \caption{}
    \end{subfigure}
    \hfill
    \begin{subfigure}{0.32\textwidth}
        \centering
        \includegraphics[width=\linewidth]{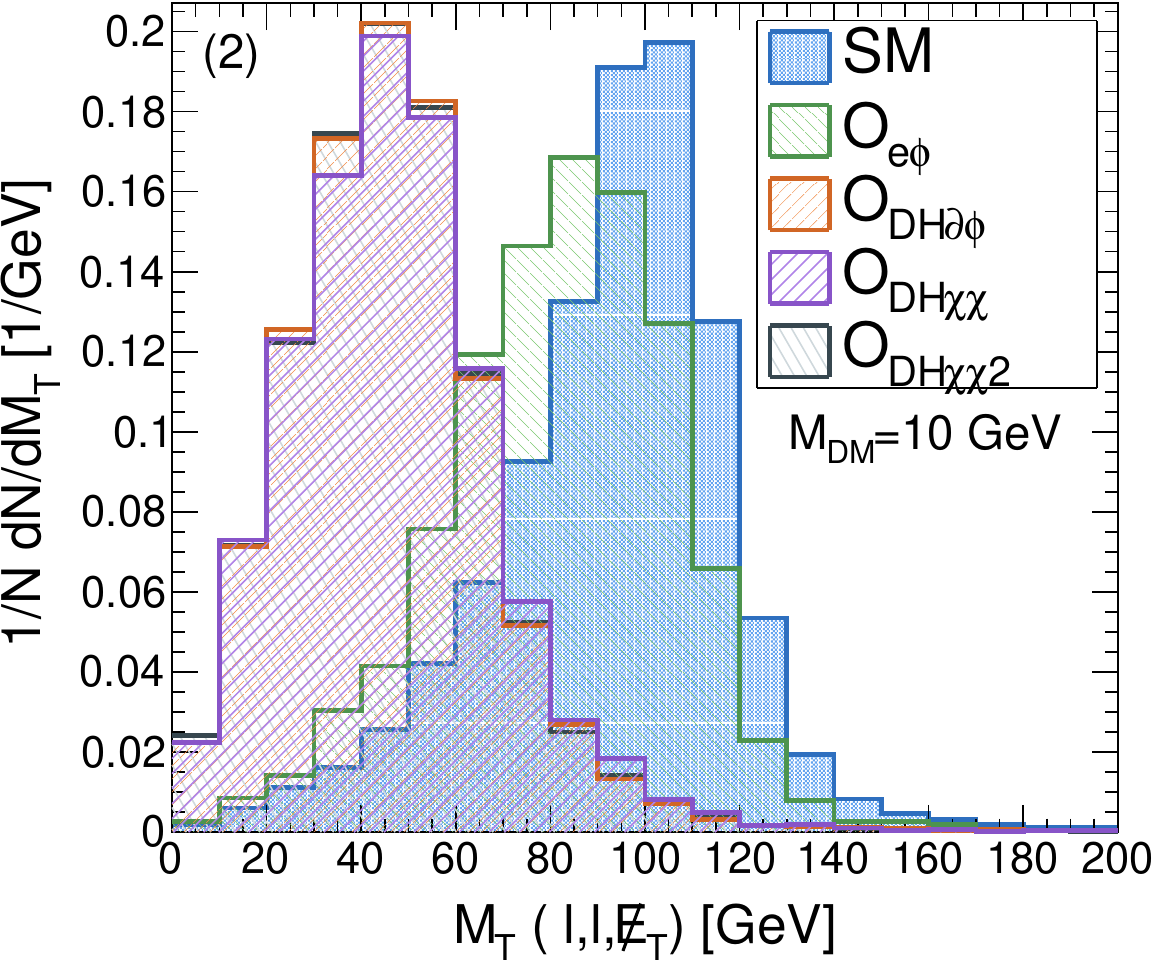}
        \caption{}
    \end{subfigure}
    \hfill
    \begin{subfigure}{0.32\textwidth}
        \centering
        \includegraphics[width=\linewidth]{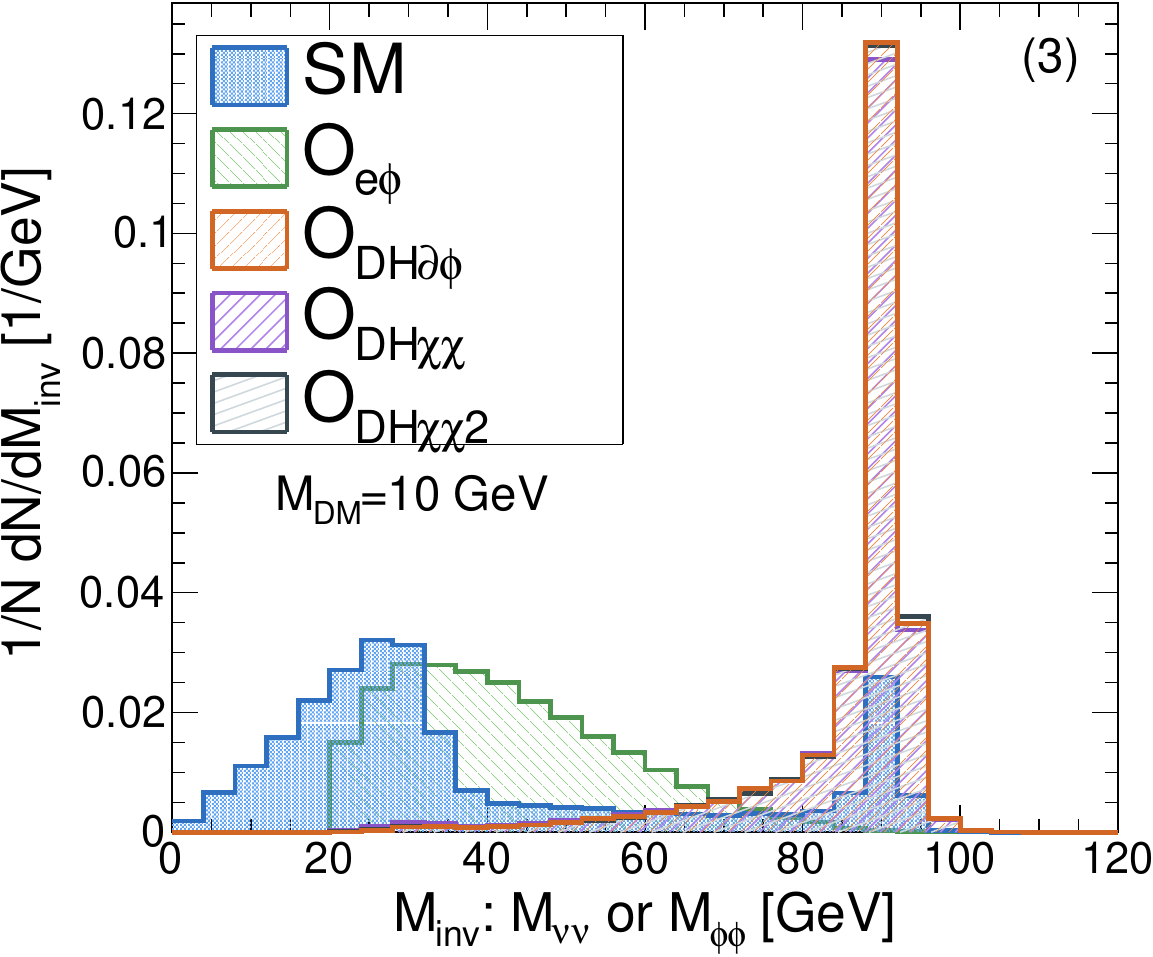}
        \caption{}
        \label{fig:dist_minv}
    \end{subfigure}
    \begin{subfigure}{0.33\textwidth}
        \centering
        \includegraphics[width=\linewidth]{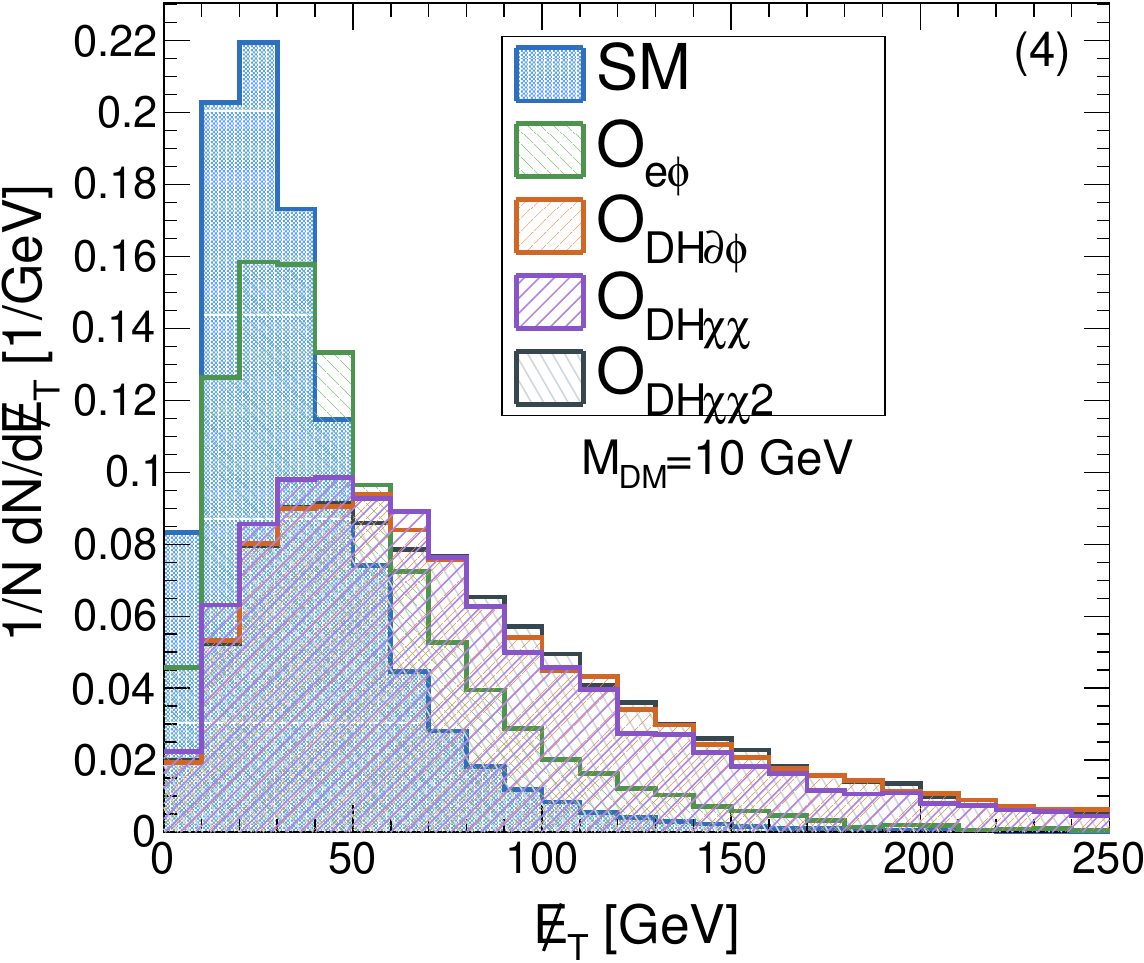}
        \caption{}
    \end{subfigure}
    \hfill
    \begin{subfigure}{0.32\textwidth}
        \centering
        \includegraphics[width=\linewidth]{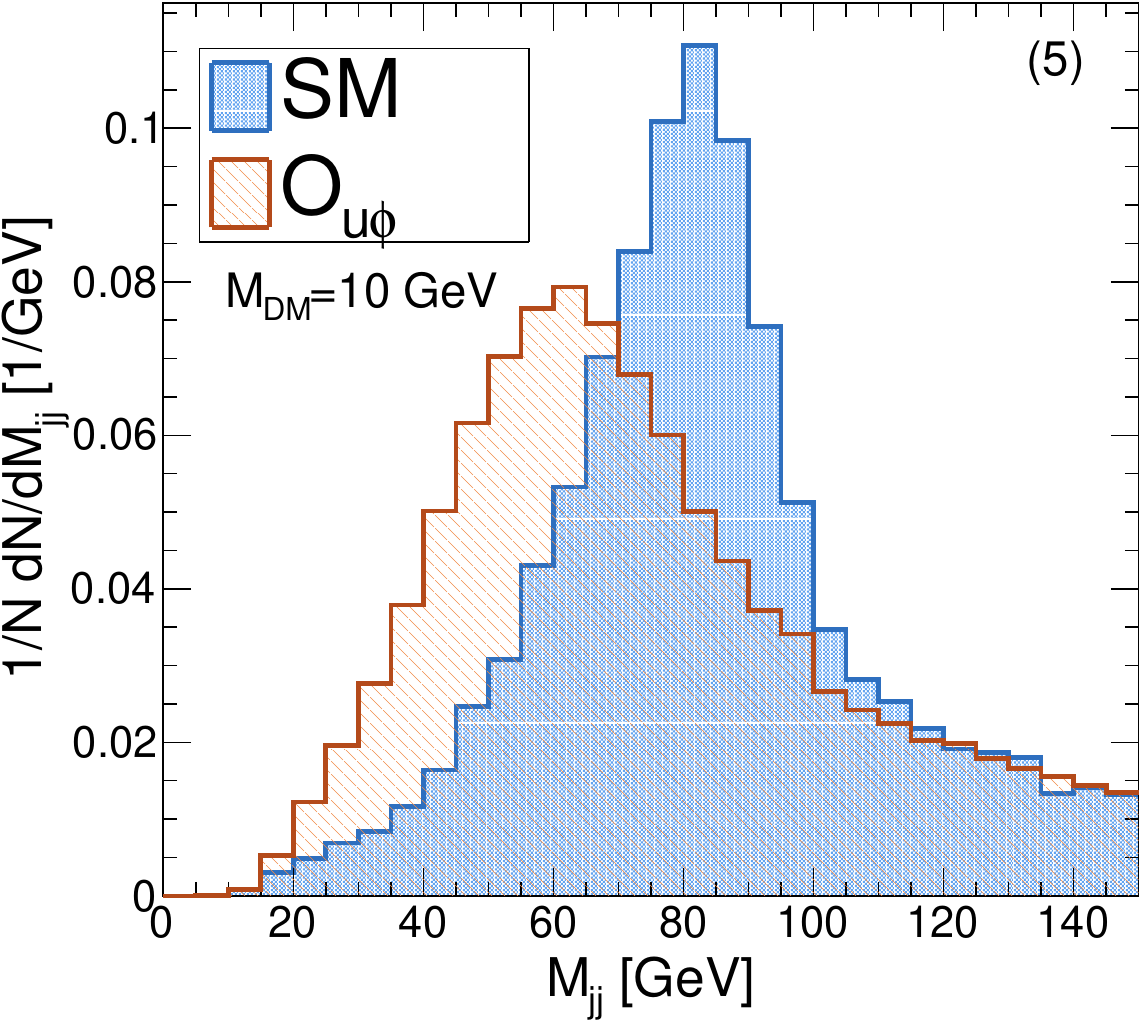}
        \caption{}
    \end{subfigure}
    \hfill
    \begin{subfigure}{0.32\textwidth}
        \centering
        \includegraphics[width=\linewidth]{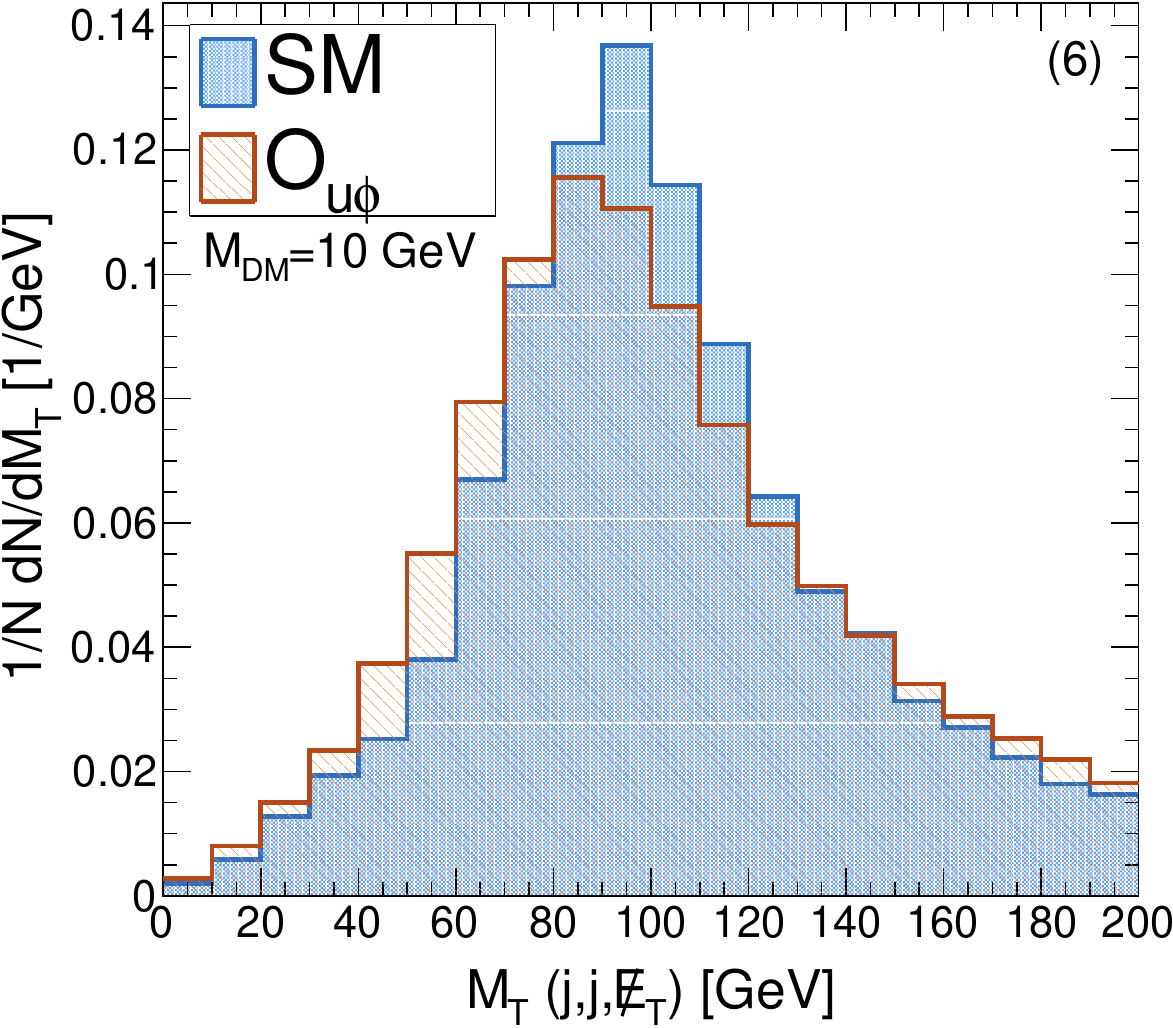}
        \caption{}
    \end{subfigure}
    \caption{\small 
    Distributions for the topology (1) in Fig.~\ref{fig:zh_cascades}: 
    (a) reconstructed invariant mass of the two leptons, 
    (b) transverse mass of the two leptons and $\MET$, 
    (c) parton-level invariant mass of the invisible system, namely 
    $\nu\bar{\nu}$ for the SM contribution through $H\to ZZ^*$ and 
    $\phi\phi$ for DSMEFT, 
    (d) $\MET$, 
    (e) invariant mass of the reconstructed jets, and 
    (f) transverse mass of the two jets and $\MET$. 
    All histograms are normalized to unit area and therefore show shape only. 
    We use $\wc=1$, $\Lambda=1~\textrm{TeV}$, and $\sqrt{s}=14~\textrm{TeV}$ in all panels.}
    \label{fig:distributions_lep}
\end{figure}

In Fig.~\ref{fig:distributions_lep}, we display the dilepton invariant mass distribution, the invisible mass distribution (which is not observable), and the transverse mass $M_T(\ell,\ell,\MET)$ distribution for the SM, as well as for all leptonic DSMEFT operators. The dilepton and invisible mass distributions for the SM show a large peak at the $Z$ mass, as expected. In these figures we use the values ${\cal C}=1,~\Lambda=1$~TeV and have taken $M_{DM}=10$~GeV as a representative value, chosen to lie in between the massless limit and the kinematic suppression that sets in for $M_{DM}\gtrsim 30$ GeV. 

For the DSMEFT operators that induce the topology in the center panel of Fig.~\ref{fig:feynmandiagrams}, $\op{DH\partial\phi}$ and~$\op{DH\chi\chi}$, there is a large $Z$ peak in the invisible mass distributions, but only a very small one in the dilepton mass distributions. 
This curious feature is due to a partial cancellation between the topologies in the first two diagrams in the bottom panel of Fig.~\ref{fig:feynmandiagrams} when the dilepton mass  $M_{\ell\ell}\sim M_Z$. The sum of these two diagrams is proportional to 
\begin{align}
   \propto  \left(1-\frac{1}{1-M_{\phi\phi}^2/M_Z^2}\right)\lesssim 0.15.
\end{align}
The numerical result for $M_{\ell\ell}\sim M_Z$ follows from energy conservation, which requires $M_{\phi\phi}<M_H-M_Z$. This has two important consequences:
\begin{itemize}
    \item The SM background $H\to ZZ^* \to \ell^+\ell^- \nu{\overline{\nu}}$ is \textit{reducible} for these DSMEFT scenarios and a cut on the dilepton invariant mass $M_{\ell\ell}$ effectively removes it, thereby eliminating the Higgs–neutrino floor.
    \item The dominant contribution to semi-visible Higgs decays for these operators is due to invisible $Z$ decays. As we show below, invisible $Z$ decays provide the main constraint in this case.
\end{itemize}

Operators that only generate the contact topology of the right panel of Fig.~\ref{fig:feynmandiagrams}, e.g., $\op{e\phi}$, do not show $Z$ peaks in either distribution. We quantify below the importance of the different SM  background processes for this case and note that semi-visible Higgs decays yield the strongest constraint on these couplings. 

\begin{figure}[t]
    \centering
    \begin{subfigure}{0.47\textwidth}
        \centering
        \includegraphics[width=\linewidth]{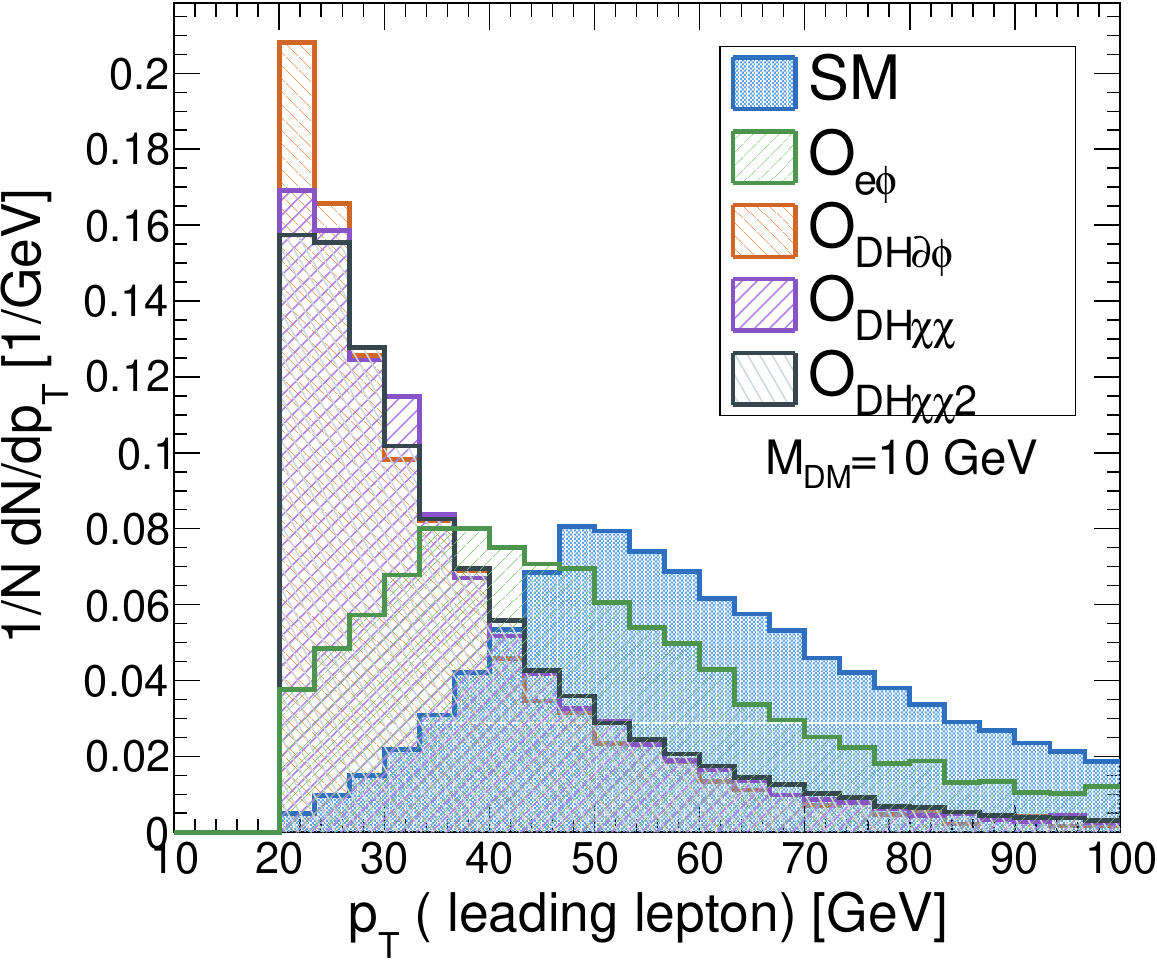}
        \caption{}
    \end{subfigure}
    \hfill
    \begin{subfigure}{0.47\textwidth}
        \centering
        \includegraphics[width=\linewidth]{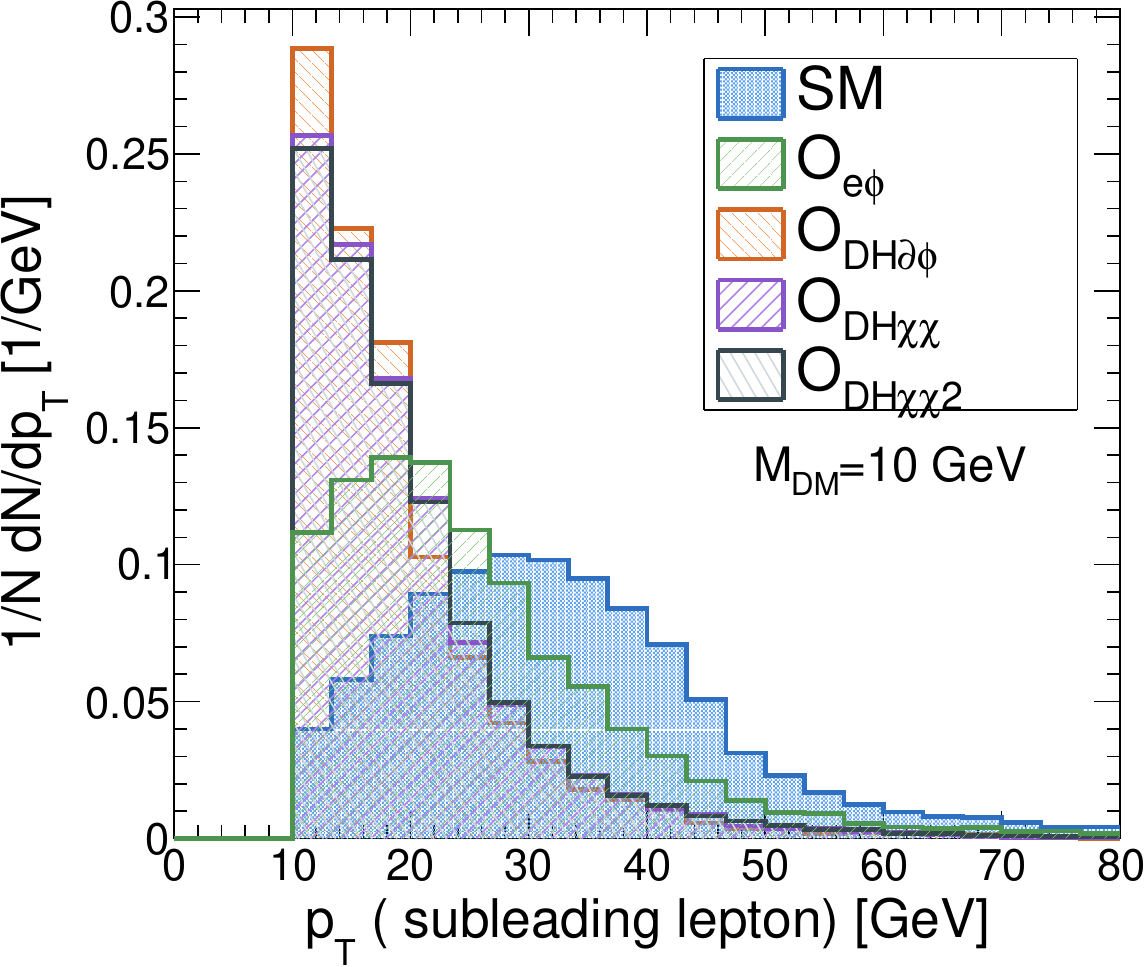}
        \caption{}
    \end{subfigure}

    \caption{\small 
    Distributions of the transverse momentum of 
    (a) the leading lepton and 
    (b) the sub-leading lepton for the SM and different DSMEFT operators. 
    All histograms are normalized to unit area and therefore show shape only. 
    We use $\wc=1$, $\Lambda=1~\textrm{TeV}$, and $\sqrt{s}=14~\textrm{TeV}$ in both panels.}
    \label{fig:distributions_lep2}
\end{figure}

In semi-visible Higgs decays, it is not possible to fully reconstruct the Higgs boson invariant mass. Instead, we reconstruct  the dilepton–$\MET$ transverse mass, \begin{align}
& \rm M_T^2(\ell,\ell,\MET)
= \big(p_{T1}+p_{T2}+\MET\big)^2
   - | \overrightarrow{p}_{T1}+\overrightarrow{p}_{T2}+\overrightarrow{p}_T^{miss}|^2
\nonumber\\
&= 2\,p_{T1}p_{T2}\big(1-\cos\Delta\phi_{12}\big)
 + 2\,p_{T1}\MET\big(1-\cos\Delta\phi_{1m}\big)
 + 2\,p_{T2}\MET\big(1-\cos\Delta\phi_{2m}\big),
\end{align}which we show in Fig.~\ref{fig:distributions_lep} (b). The ${\rm M_T}(\ell,\ell,\MET)$ spectrum is kinematically bounded above by the Higgs mass $M_H \simeq 125~\text{GeV}$; this endpoint can be exploited to suppress backgrounds with similar final states that do not originate from a Higgs boson decay. For the derivative DSMEFT operators $\opdhphi$ and $\opdhchi$, the distribution shifts to lower values since the lepton $p_T$ distribution peaks at much lower $p_T$, as can be seen in Fig.~\ref{fig:distributions_lep2}. 

For the hadronic semi-visible Higgs decay modes, we present the invariant mass of the two reconstructed jets, and the transverse mass of the two jets and $\MET$ in Fig.~\ref{fig:distributions_lep}~(e) and Fig.~\ref{fig:distributions_lep}~(f), respectively. The broadening of the peaks relative to those in leptonic final states is expected due to ambiguities in the jet reconstruction.

Although both channels yield $2j+2\ell+E_T^{\rm miss}$, the signal is essentially channel-specific: lepton-only operators populate the leptonic category and quark operators the hadronic category, with only small spillover ($0.7\%$ and $2\%$, respectively). For SM backgrounds, the cuts $m_{\ell\ell}<60$ GeV and $m_{jj}<60$ GeV suppress events where the corresponding pair comes from an on-shell $Z$, making the contribution to the other channel negligible.

\subsection{Signal and SM backgrounds}

In addition to SM $ZH$ production followed by the decay chain $H\to ZZ^* \to \ell^+\ell^- \nu{\overline\nu}$, several other SM processes are potential sources of background to the semi-visible Higgs decays. The leading order estimates for these background cross-sections are presented in Table~\ref{tab:xs_summary} (left panel), whereas representative values for the signal cross-sections for different operator contributions are presented in Table~\ref{tab:xs_summary} (right panel). These background cross-sections are significantly larger than the signal, and we now turn our attention to ways of suppressing them. QCD multijet backgrounds are negligible in this channel after requiring two isolated leptons and sizable $\MET$~\cite{CMS:2023vzh,ATLAS:2016vox}, so we do not include them.

\begin{table}
\captionsetup{justification=raggedright,singlelinecheck=false}
\centering
  \begin{subtable}[b]{.35\linewidth}
    \centering
    \begin{tabular}{|c|c|}
    \hline
      Process & ~$\sigma_{\text{SM,LO}}$ [fb]~ \\
      \hline
      \small{ZH-lep: $H\to ZZ^*\to\ell^+\ell^- +\nu{\overline{\nu}}, Z\to jj$} & 0.05 \\
      \small{ZH-had: $H\to ZZ^*\to 2 j +\nu{\overline{\nu}}, Z\to \ell^+\ell^-$} & 0.05 \\
      $ZZ$, $Z\to jj$, $Z\to \ell^+\ell^-$ & 547 \\
      $ZZZ$, $Z\to jj$, $Z\to \ell^+\ell^-$, $Z\to \nu{\overline{\nu}}$ & 0.33 \\
      $WWZ$, $ \to jj\,\ell^+\,\ell^+\,\nu\,{\overline{\nu}}$ & 2.2 \\
      $t\bar t Z$, $ \to jj\,\ell^+\,\ell^-\,\nu\,{\overline{\nu}} \,b\,\bar{b}$ & 17.7 \\
      $tWZ$, $ \to jj\,\ell^+\,\ell^-\,\nu\,{\overline{\nu}}\, b$ & 6.7 \\
      \small{$\rm ZH$-W: $H\to WW^*\to\ell^+\ell^- +\nu{\overline{\nu}}, Z\to jj$} & 0.06 \\
      \small{$t\bar{t}$-inc: $t\to b\,\ell^+\nu,\; \bar{t}\to \bar{b}\,\ell^-\bar{\nu}$} & 45719 \\
      \hline
    \end{tabular}
  \end{subtable}\hfill
  \begin{subtable}[b]{.38\linewidth}
    \centering
    \begin{tabular}{|c|c|c|}
    \hline
      WC & $~\small{{C\over \Lambda^2}~[{\textrm{TeV}}^{-2}]}$~ & ~$\rm \delta\sigma_{LO}$ [fb]~ \\
      \hline
      $\wcdhphi$ & 10 & 0.005 \\
      $\wcephi$  & 10 & 0.004 \\
      $\wcuphi$  & 10 & 0.001 \\
      $\wcdphi$  & 10 & 0.001 \\
      $\wcdhchi$ & 10 & 0.021 \\
      $\wc_{DH\chi\chi2}$ & 10 & 0.019 \\
      \hline
    \end{tabular}
  \end{subtable}
  \captionsetup{justification=raggedright,singlelinecheck=false}
  \caption{Leading-order cross-sections at $\sqrt{s}=14$ TeV at the LHC with selections
  $p_T(j)>20$ GeV, $p_T(\ell_1)>20~\text{GeV}$, and $p_T(\ell_2)>10~\text{GeV}$.
  Left: SM backgrounds (including the relevant branching fractions); for $t\bar t Z$, $tWZ$, and $WWZ$, the quoted rates are inclusive sums over all decay combinations that can yield the reconstructed $2$--$3\ell+jj+\MET$ final state, allowing for the possibility that in the $3\ell$ case one lepton is missed.
 Right: Contributions to total cross section (with cuts) from DSMEFT operators. The total cross section is 
$\sigma_{TOT,LO}=$ $\sigma_{SM,LO}+\Sigma {C_i^2\over \Lambda^4}f_{ii}=\sigma_{SM,LO}+\delta \sigma_{LO}$, since there is no interference between the SM and the DSMEFT contributions.}
  \label{tab:xs_summary}
\end{table}

\begin{figure}[t]
    \centering
    \begin{subfigure}{0.49\textwidth}
        \centering
        \includegraphics[width=\linewidth]{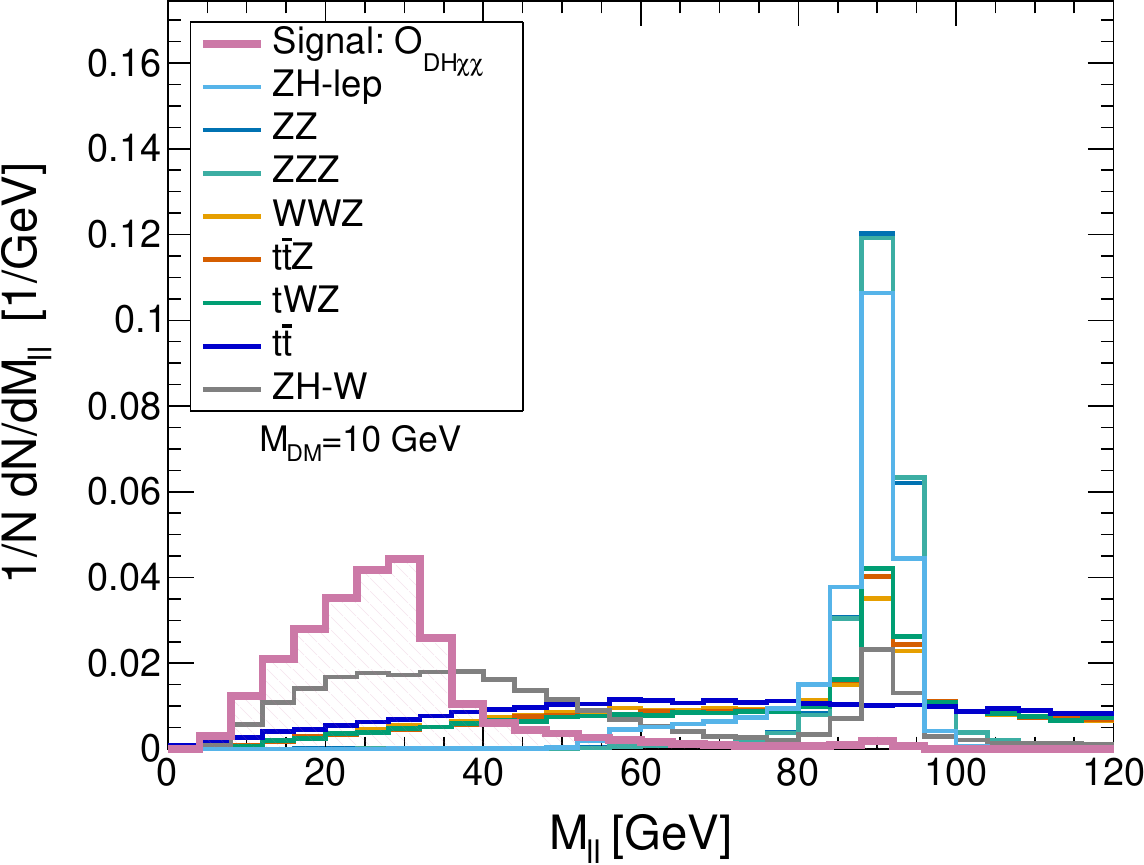}
        \caption{}
    \end{subfigure}
    \hfill
    \begin{subfigure}{0.47\textwidth}
        \centering
        \includegraphics[width=\linewidth]{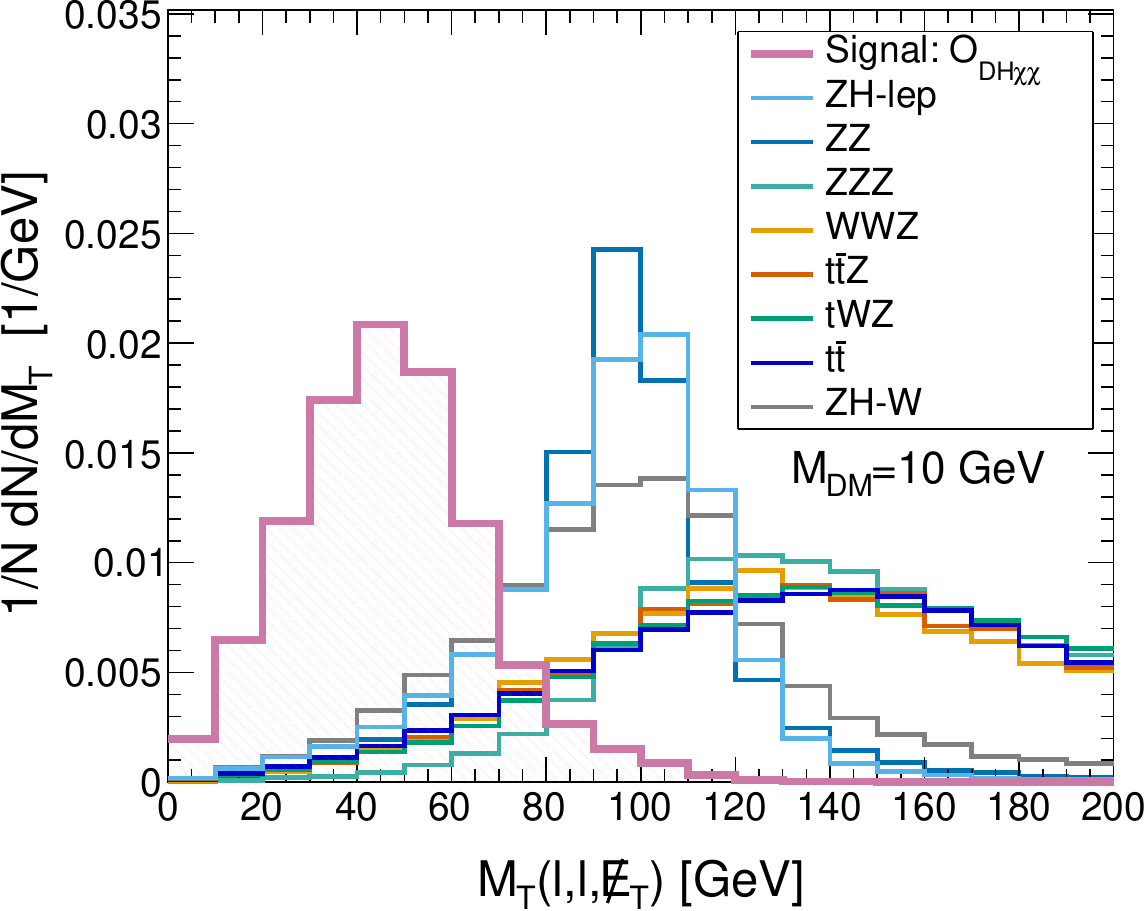}
        \caption{}
    \end{subfigure}
    \begin{subfigure}{0.47\textwidth}
        \centering
        \includegraphics[width=\linewidth]{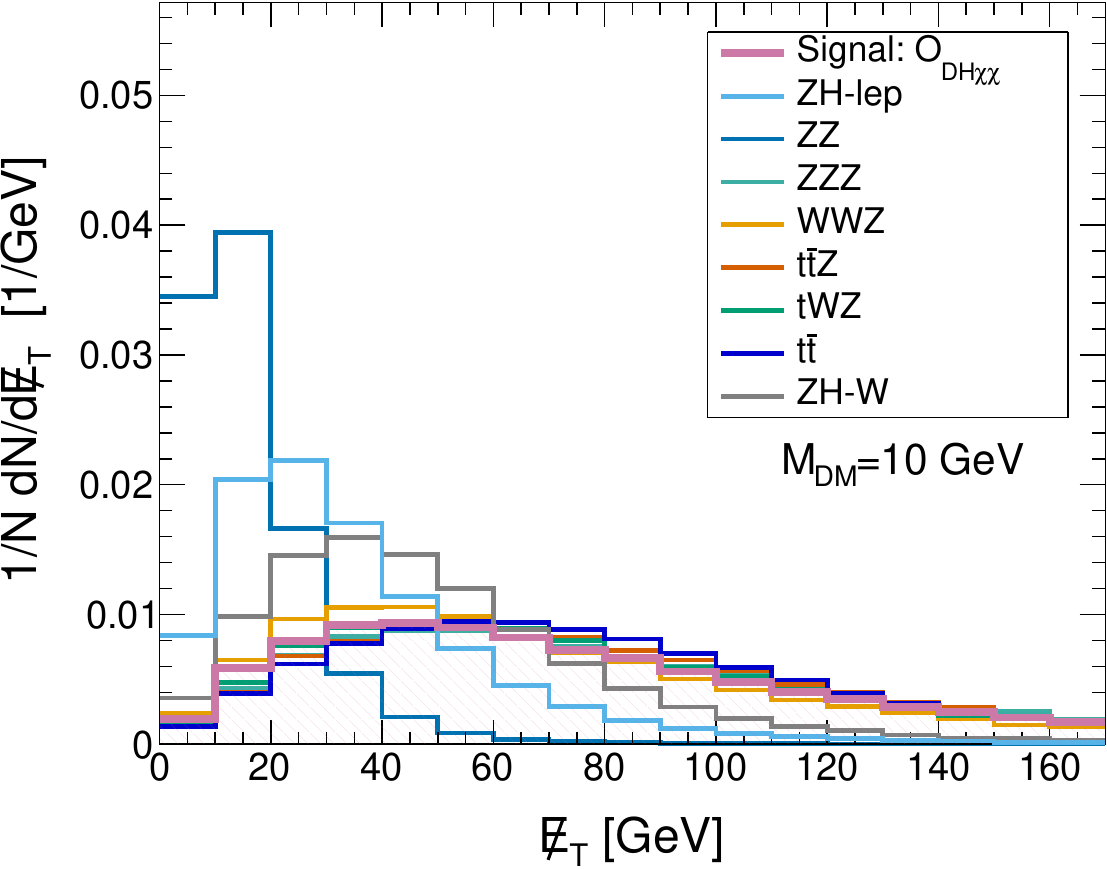}
        \caption{}
    \end{subfigure}
    \hfill
    \begin{subfigure}{0.47\textwidth}
        \centering
        \includegraphics[width=\linewidth]{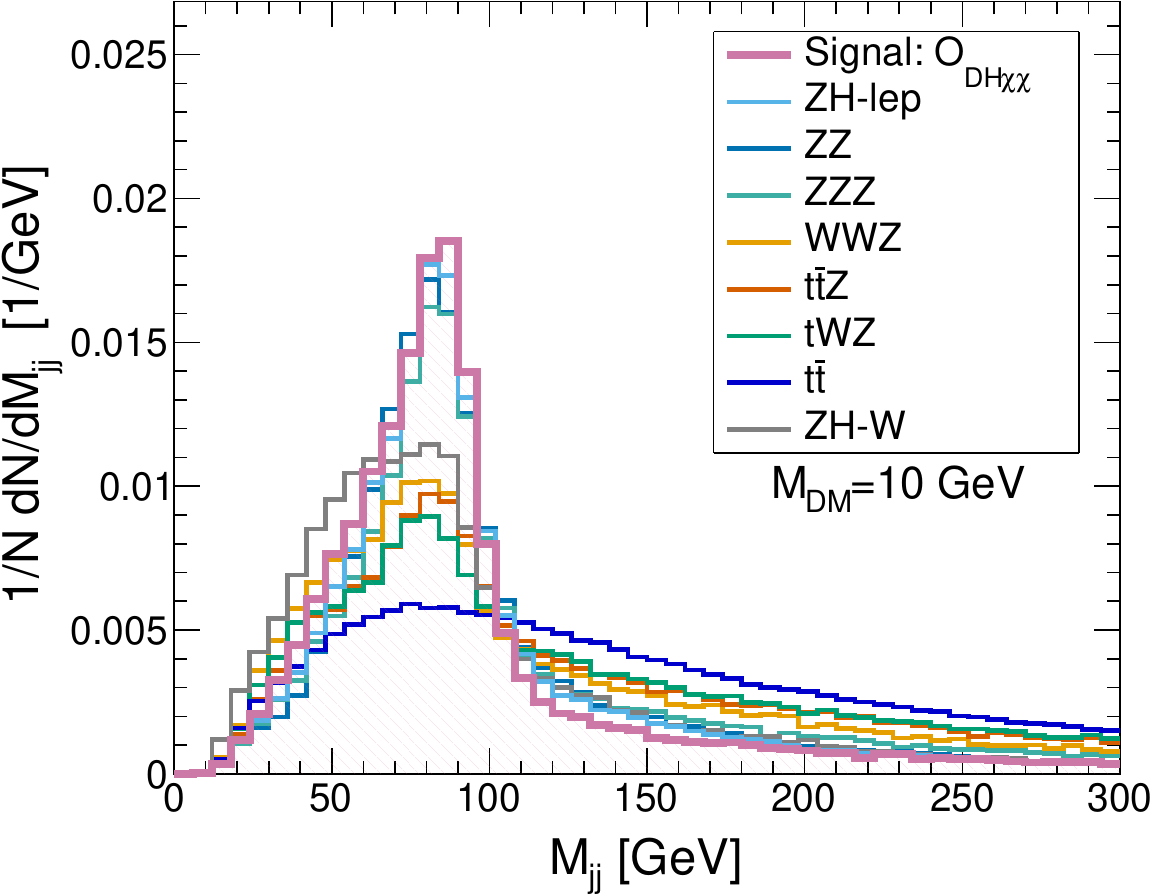}
        \caption{}
    \end{subfigure}
    \caption{\small 
    Distributions of 
    (a) the dilepton invariant mass, 
    (b) the transverse mass of the two leptons and $\MET$, 
    (c) $\MET$, and 
    (d) the invariant mass of the two leading non-$b$ jets. 
    All histograms are normalized to unit area and therefore show shape only. 
    We use $\wc=1$, $\Lambda=1~\textrm{TeV}$, and 
    $\sqrt{s}=14~\textrm{TeV}$ in all cases.}
    \label{fig:distributions_lep3}
\end{figure}

We begin with a simple cut-based analysis to understand the effectiveness of various kinematic variables in reducing the background for the contributions from the operators $\opdhchi,~\op{DH\chi\chi2}$, $\opdhphi$, and ~$\opephi$ with $\rm M_{DM}=10 ~ GeV$ and ${C\over\Lambda^2}=10\;\rm TeV^{-2}$ as benchmark points. In Fig.~\ref {fig:distributions_lep3} , we compare several distributions for different background channels at the reconstructed level and compare them with the signal distributions for $\opdhchi$, while the other signal distributions can be found in Fig.~\ref {fig:distributions_lep}. All of these backgrounds have a peak at the  Z-boson mass, which is dominant for the ZH-lep, ZZ, ZZZ, and WWZ channels. An  $\rm M_{\ell\ell}<60~GeV$ selection cut reduces these backgrounds significantly.

A selection cut of  $\rm M_T(\ell,\ell,\MET)<125~ GeV$ significantly reduces the backgrounds from $ZZZ,WWZ$, $t\bar{t}Z,tWZ$ as shown in Fig.~\ref {fig:distributions_lep3} (b). A selection cut of $\rm \MET>20.0\; GeV$ turns out to be helpful for reducing the $ZZ^*$ background and to some extent for the $ZH$-lep background. Finally, the top-quark background includes b-jets in the final state and can be suppressed by vetoing these jets. An example cut-flow based analysis is presented in Table~\ref{tab:cut_flow}. For the leptonic semi-visible Higgs decay, where the other Z decays hadronically, we impose an additional Z-tag requirement $70< M_{jj} <110~\mathrm{GeV}$ which further suppresses the $t\bar{t}$ background.

\begin{table}[t]
  \centering
  \caption{Effects of various selection criteria on signals and SM backgrounds, with $M_{DM}=10$ GeV,  and $\Lambda=1~{\textrm{TeV}}$.  Note that the C's are dimensionless.  The rows show the cumulative effects of applying the various cuts.}
  \begin{subtable}{\textwidth}
    \centering
    \caption{Signals: $\wc_{DH\chi\chi},\wc_{DH\chi\chi2},\wc_{DH\partial\phi},\wc_{e\phi}$.}
    \label{tab:cut_flow_signals_sub}
    \vspace{0.4em}
      \begin{tabular}{|c|c|c|c|c|}
      	\hline
      	& $\wc_{DH\chi\chi}=10$
      	& $\wc_{DH\chi\chi2}=10$ & $\wc_{DH\partial\phi}=10$ & $\wc_{e\phi}=10$ \\
      	\hline
      	Cross-section (fb)  & $0.021$ & $0.019$  & $0.005$ & $0.004$ \\
      	\hline
      	$\rm M_{\ell\ell}<60~GeV$  
      	& {$0.013$} & {$0.012$}  & {$0.0026$} & {$0.0011$} \\
      	\hline
      	$\rm M_T(\ell,\ell,\MET)<125~ GeV$  
      	& {$0.013$} & {$0.012$}  & {$0.0026$} & {$0.0011$} \\
      	\hline
      	$\MET>20.0\; GeV$   
      	& {$0.012$} & {$0.011$}  & {$0.0023$} & {$0.0009$} \\
        \hline $\rm 70< M_{jj} <110~\mathrm{GeV}+N(\cancel{b}-jet)=2$  & {$0.004$}  & {$0.004$}  & {$0.001$}  & {$0.0003$}\\ 
        \hline
      	$\rm b\text{-}jet~veto$  
      	& {$0.004$}  & {$0.004$} & {$0.001$} & {$0.0003$} \\
      	\hline
      \end{tabular}
  \end{subtable}

  \vspace{1em}

  \begin{subtable}{\textwidth}
    \centering
    \caption{Backgrounds: ZH-lep, ZH-W, ZZ, ZZZ, WWZ, $\rm t\bar{t}Z$, tWZ, $t\bar{t}$-inc.}
    \label{tab:cut_flow_backgrounds_sub}
    \vspace{0.4em}
    \resizebox{\linewidth}{!}{
      \begin{tabular}{|c|c|c|c|c|c|c|c|c|}
      	\hline   & ZH-lep  & ~~ZH-W~~ & ~~~ZZ~~~  & ZZZ  & ~~WWZ~~  & ~~$\rm t\bar{t}Z$~~  & ~~tWZ~~ & $t\bar{t}$-inc  \\
      	\hline Cross-section (fb)  & $0.05$  & $0.06$  & $547$  & $0.33$  & $2.2$  & $17.7$  & $6.7$ & $45719$\\ 
      	\hline $\rm M_{\ell\ell}<60~GeV$  & {$1.35 \times 10^{-3}$}  & {$0.014$}  & {$5.87$}  & {$4.27 \times 10^{-3}$}  & {$0.15$}  & {$2.4$}  & {$0.7$} & $3533$\\ 
      	\hline $\rm M_T(\ell,\ell,\MET)<125~ GeV$  & {$1.33 \times 10^{-3}$}  & {$0.013$}  & {$5.76$}  & {$2.30 \times 10^{-3}$}  & {$0.097$}  & {$1.4$}  & {$0.4$} & $2025$\\ 
      	\hline $\rm \MET>20.0\; GeV$   & {$9.7 \times 10^{-4}$}  & {$0.011$}  & {$1.12$}  & {$1.89 \times 10^{-3}$}  & {$0.085$}  & {$1.2$}  & {$0.36$} & $1782$\\ 
        \hline $\rm 70< M_{jj} <110~\mathrm{GeV}+N(\cancel{b}-jet)=2$   & {$2.3 \times 10^{-4}$}  & {$0.002$}  & {$0.09$}  & {$2 \times 10^{-4}$}  & {$0.016$}  & {$0.13$}  & {$0.033$} &  74\\ 
      	\hline $\rm b-jet ~veto$  & {$2.2 \times 10^{-4}$}  & {$0.002$}  & {$0.09$}  & {$2 \times 10^{-4}$}  & {$0.016$}  & {$0.0065$}  & {$0.006$}  & 1.4\\ 
      	\hline
      \end{tabular}}
  \end{subtable}
\label{tab:cut_flow}

\end{table}

 The total background cross-section after cuts is $\sigma_{bkg}=1.5~\rm fb$. For the illustrative choice of ${C\over \Lambda^2}=10\text{ TeV}^{-2}$ and $M_{DM}= 10\;\rm GeV$ used in Table~\ref{tab:cut_flow_signals_sub}, we find the signal sensitivities to $\wcdhchi$ and $\wc_{DH\chi\chi2}$ are around $\mathbb{S}_0=S/\sqrt{S+B}=0.18 \,\sigma$, $\mathbb{S}_0=0.045 \,\sigma$ for $\wcdhphi$, and lowest for $\wcephi$ with $\mathbb{S}_0=0.013 \,\sigma$. As there is no DSMEFT-SM interference, the DSMEFT contribution is purely quadratic and kinematic shapes also remain unchanged over the range of $C$ values considered here. So we can rescale the individual coefficients by,
\begin{align}
    C^{min}=C_{0}\times \left(\frac{\mathbb{S}^{\rm target}}{\mathbb{S}_0}\right)^{1/2},
    \label{eq:scaling}
\end{align}
to find the minimum values of $C^{min}$ required to obtain the target sensitivity $\mathbb{S}^{\rm target}$, given an initial value $C_0$. This indicates that the minimum values to achieve a $3\sigma$ signal significance is around $C^{min}\simeq 40~\rm TeV^{-2}$ for ${\wcdhchi}$ and $\wc_{DH\chi\chi2}$, $C^{min}\simeq 81~\rm TeV^{-2}$ for $\wcdhphi$, and $C^{min}\simeq 152~\rm TeV^{-2}$ for $\wcephi$ with a cut-based analysis.

A certain amount of improvement is expected through a multivariate analysis (MVA), which takes into account the possible correlations among variables and optimizes the selection criteria through a multi-dimensional variable space. We employ a boosted decision tree (BDT) with gradient boosting and decorrelation within the standard ROOT TMVA framework~\cite{Hocker:2007ht,Voss:2007jxm}. This method builds a sequence of disjoint decision trees by optimizing cuts on the training kinematic variables. After each tree, events misclassified as signal or background are given higher weights, and a new tree is trained on this reweighted sample. Iterating this procedure progressively refines the classifier and finally yields an optimized discriminator. In addition to the kinematic variables described above, we also use various $\Delta\phi$ distributions, such as $\Delta \phi (\ell,\ell)$, $\Delta \phi (\ell_1,\MET)$, $\Delta \phi (\ell_2,\MET)$, as well as the  $\rm p_T$  of the leading lepton, the invariant mass of the two leading jets, M(jet1,jet2), for the hadronically decaying Z-boson in the signal (see Fig.~\ref{fig:zh_cascades}), the invariant mass $M(\ell,\ell,j,j)$,  the minimum between the transverse mass combinations of the $\MET$  and the two leptons ($\rm M_T^{min}(\ell, \MET)$), $R_T(2)=(p_T(j_1)+p_T(j_2))/\sum_i p_T(j_i)$~\cite{Guchait:2011fb}, the number of non-b-jets with $p_T>20$ GeV, and the resultant $p_T$ of the two leptons. These variables are presented in Table~\ref{tab:variable_ranking} according to their relative power of separation or effectiveness in background reduction in decreasing order. We use 60\% of the events from each signal and background sample for training the BDT, reserving the remaining 40\% for testing. The final BDT classifier is then used as a single discriminant to separate signal from background.

\begin{figure}
    \centering
    \includegraphics[width=0.49\linewidth]{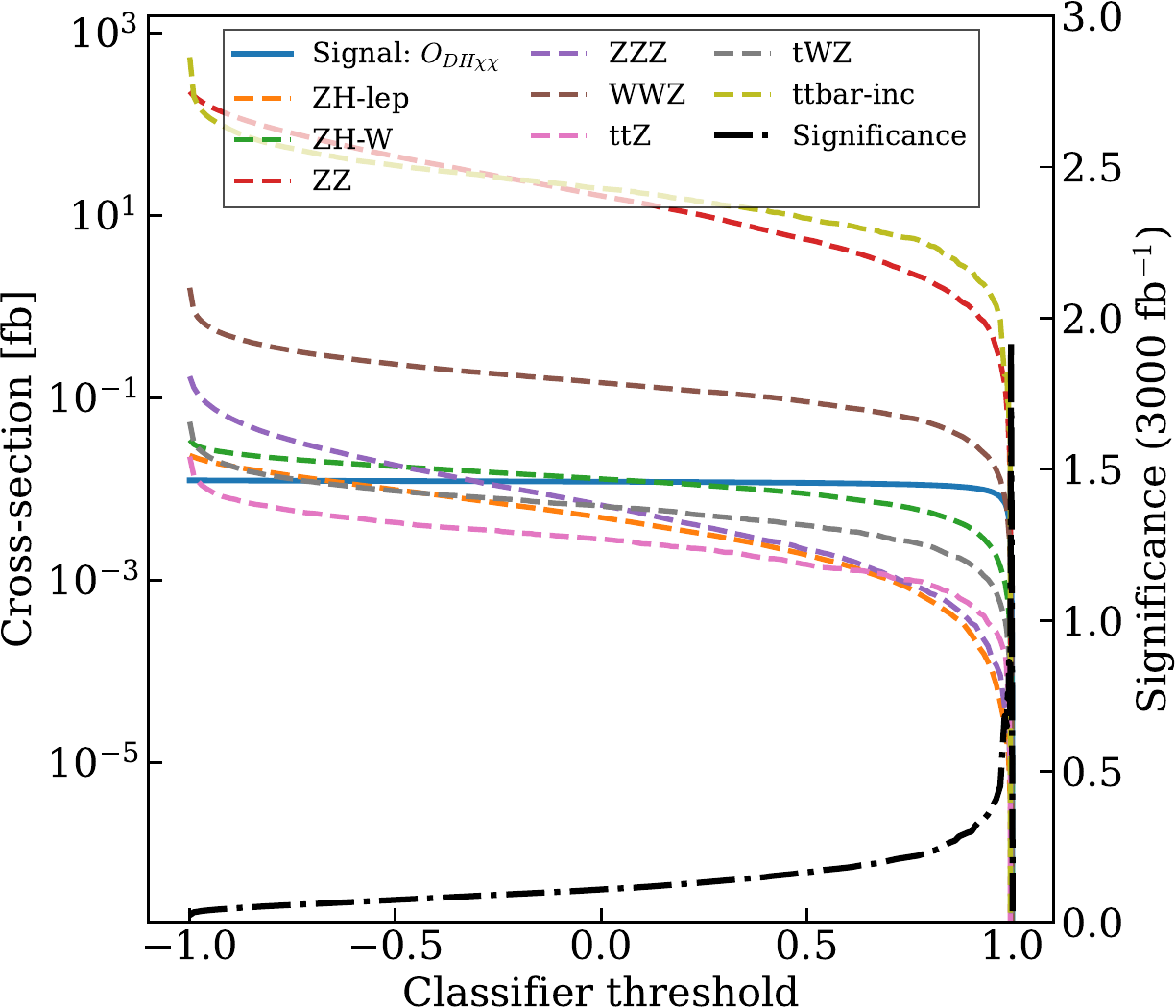}
     \includegraphics[width=0.48\linewidth]{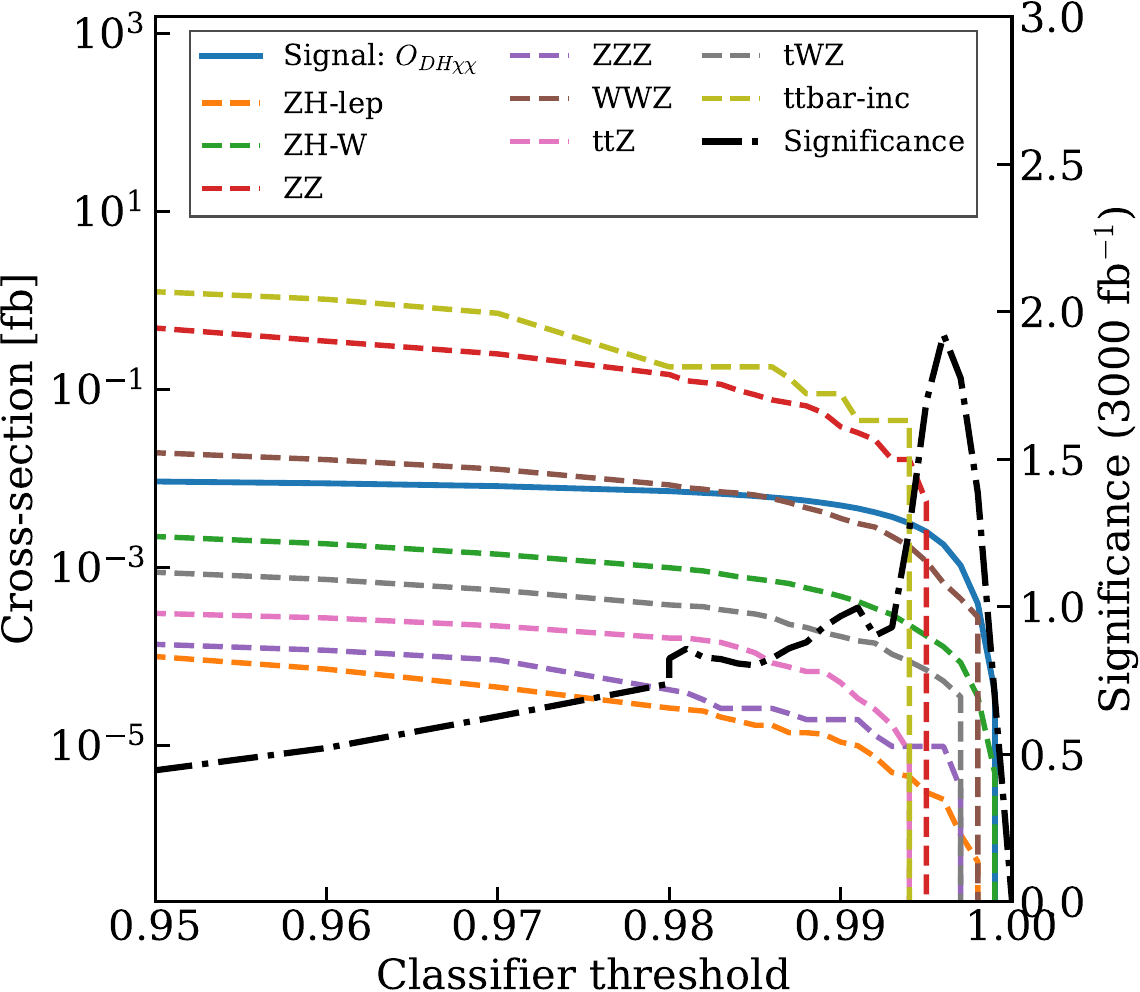}
    \caption{Distributions of the effective cross-sections and signal significance at the LHC, with $\sqrt{s}=14$ TeV and $\mathcal{L}=3000~\text{fb}^{-1}$ as functions of the MVA discriminator for the fermionic operator scenario $\opdhchi$ (${\wcdhchi\over \Lambda^2}=10~\text{TeV}^{-2}$ and  $\rm M_{DM}=10~\rm GeV$), where we train the MVA with signal and background events after b-jet vetoing and requiring two leptons. The right panel shows a zoomed-in view of the left panel, highlighting the region with larger classifier-threshold values.}
    \label{fig:BDT_sensitivity}
\end{figure}

 As an illustration, we show the effective cross sections for the fermionic DM scenario with the operator $\opdhchi$, together with the corresponding SM backgrounds, after applying a cut on the BDT (MVA) output classifier in Fig.~\ref{fig:BDT_sensitivity}. For each choice of the BDT threshold, we also report the resulting expected significance at an integrated luminosity of $\mathcal{L}=3000~\text{fb}^{-1}$. The scan over thresholds highlights the usual trade-off between background rejection and signal efficiency and identifies the working point that maximizes the expected sensitivity. Applying a classifier cut at 0.997, we observe a significant improvement in signal significance ($S/\sqrt{S+B}$) relative to the simple cut-based analysis presented above.\footnote{But note that BDT includes more variables than our cut-based analysis.} 
\begin{table}
	\centering
	\begin{tabular}{|c|c|c|}
		\hline
		\textbf{~~Rank~~} & \textbf{~~Variable~~} & \textbf{Description} \\
		\hline
        1 & $\rm M_T(\ell,\ell, \MET)$           & Transverse mass of two leptons and $\MET$ \\
		2 & $\rm M_{\ell\ell}$   & Invariant mass of two leptons \\
        3 & $\rm M_T^{min}(\ell, \MET)$ & Minimum of the transverse masses $\rm M_T(\ell_1, \MET)$ and $\rm M_T(\ell_2, \MET)$ \\
        4 & $\textrm{M}(\ell, \ell, j, j)$ & Invariant mass of the $\ell_1$, $\ell_2$, $j_1$ and $j_2$\\
		5 & $\rm p_T (\ell)$             & Transverse momentum of the leading lepton \\
		7 & $\Delta \phi (\ell_1,\MET)$       & $\Delta\phi$ between leading lepton and $\MET$ \\
        8 & $\rm p_T(\ell_1)$      & $\rm p_T$ of the leading leptons  \\
        9 & $R_T(2)$  & Ratio of $p_T(j_1)+p_T(j_2)$ and HT (scalar sum of $p_T$ of all jets)  \\
        10 & $\rm N(\cancel{b}-jets)$         &Number of non-b-jets with $\rm p_T>20~GeV$\\
        11 & $\rm M(jet1,jet2)$      & Invariant mass of two jets \\
        12 & $\Delta \phi (\ell,\ell)$          & $\Delta \phi$ between two leptons \\
        13 & $\rm p_T(\ell_1,\ell_2)$      & Resultant $\rm p_T$ of the two leptons  \\
        14 & $\Delta \phi (\ell_2,\MET)$       & $\Delta\phi$ between leading lepton and $\MET$ \\
        15 & $\rm \MET$           & Missing $\rm E_T$ \\
		\hline
	\end{tabular}
	\caption{ Ranking of input variables for the leptonic semi-visible Higgs decays shown in Fig.~\ref{fig:zh_cascades} (left) corresponding to the $\opdhchi$-scenario (${\wcdhchi\over \Lambda^2}=10~\text{TeV}^{-2}$ and  $\rm M_{DM}=10~\rm GeV$) by relative separation power. The higher the rank, the more discriminating power.
\label{tab:variable_ranking}}
\end{table}

Finally, we observe that the signal characteristics vary depending on the choice of operators and DM masses. To separately treat different kinematic regions, we split the signal into three regions based on the DM mass: (a) $\rm M_{DM}:[1,20]~ GeV$, (b)  $\rm M_{DM}:[20,40] ~GeV$, (c)  $\rm M_{DM}:[40,50] ~GeV$. For each DSMEFT operator, we perform the MVA separately for these three kinematic regions by training the BDT at the mean values of $\rm M_{DM}$ corresponding to each region. Using the BDT that is trained at the nearest masses, we first calculate the sensitivity to each of the operators with different DM-mass choices for a given value of the coefficient, and then use the scaling of Eq.~\ref{eq:scaling} to calculate the lower limit of each coefficient that can be probed at the HL-HLC with $3000~{\rm{fb}}^{-1}$ at the $3\sigma$ level. The results are presented in Fig.~\ref{fig:bounds_leptonic}.

\subsection{Operator categories and kinematic discrimination}
 
\begin{figure}[t]
    \centering
    \begin{subfigure}{0.47\textwidth}
        \centering
        \includegraphics[width=\linewidth]{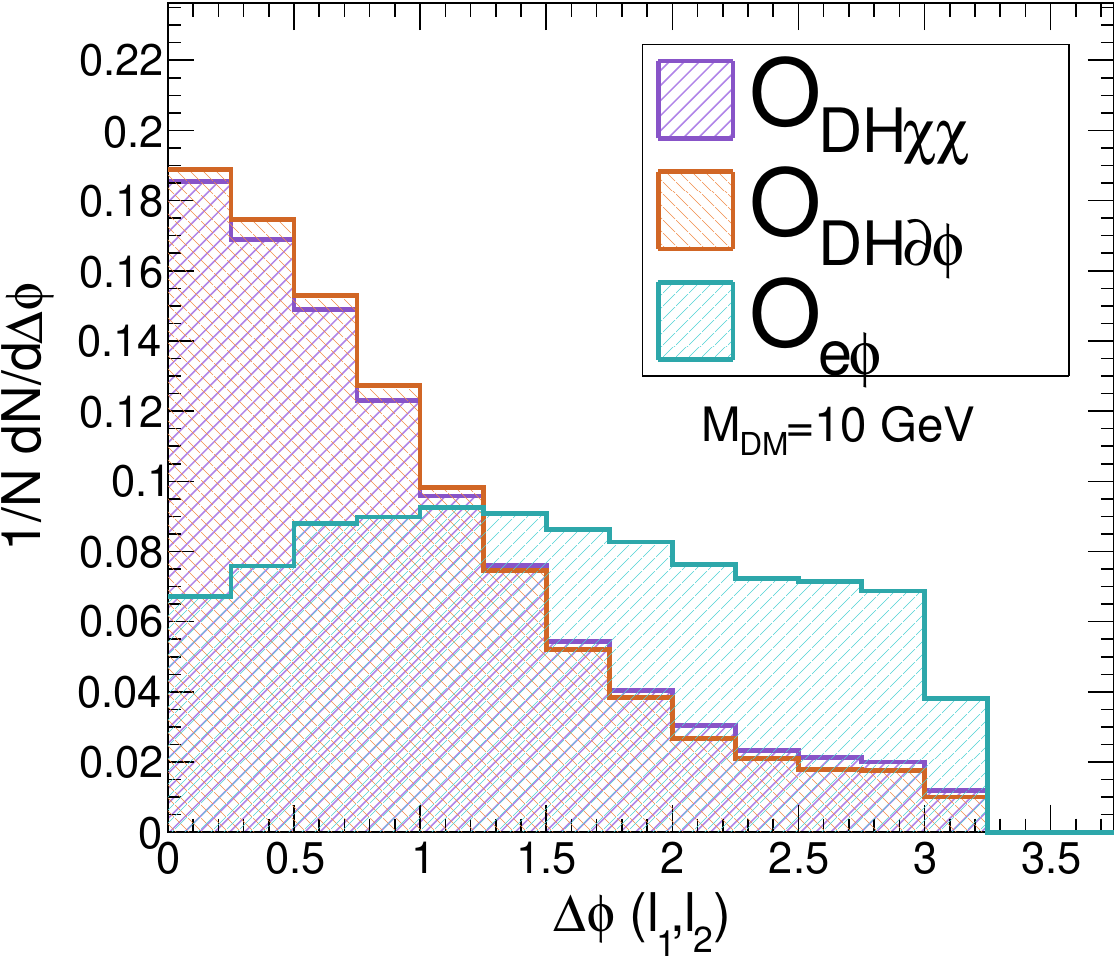}
        \caption{}
    \end{subfigure}
    \hfill
    \begin{subfigure}{0.47\textwidth}
        \centering
        \includegraphics[width=\linewidth]{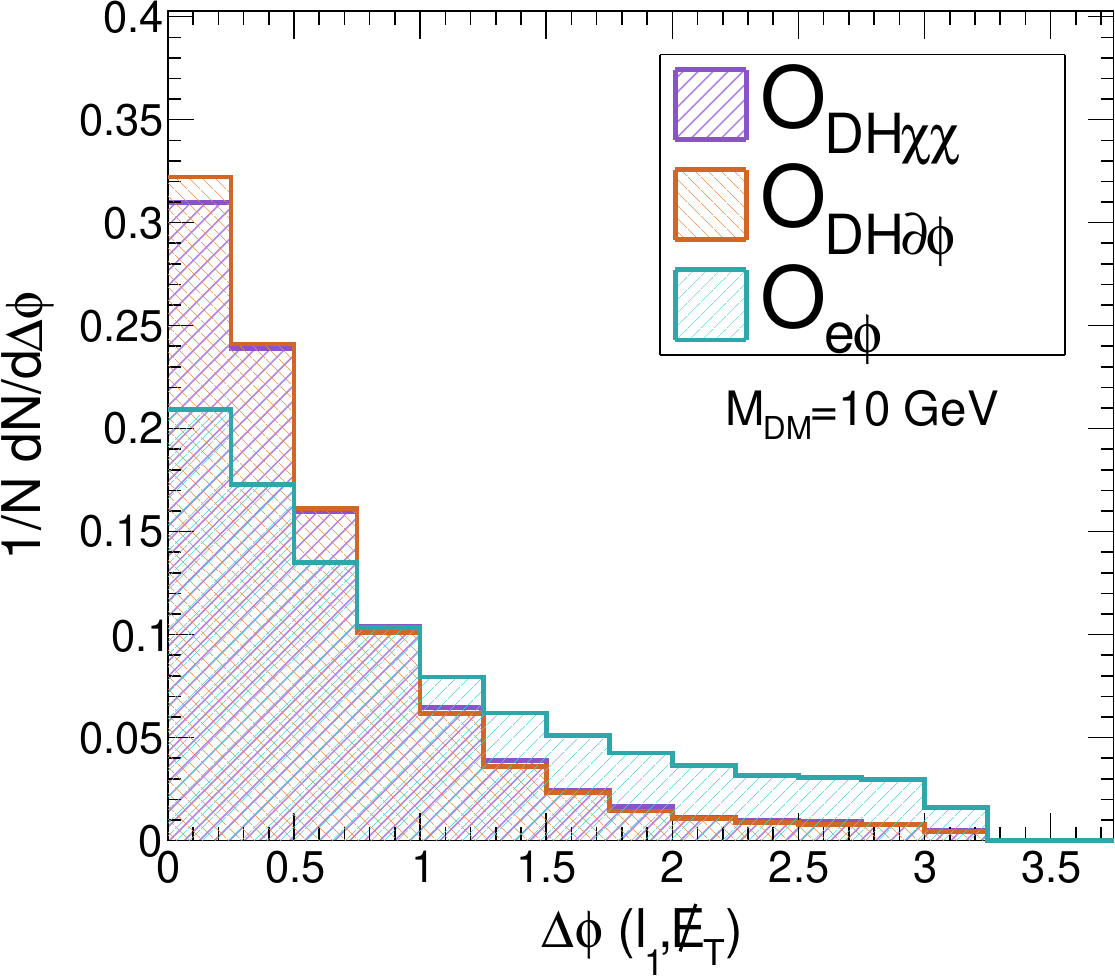}
        \caption{}
    \end{subfigure}
    \hfill
    \begin{subfigure}{0.47\textwidth}
        \centering
        \includegraphics[width=\linewidth]{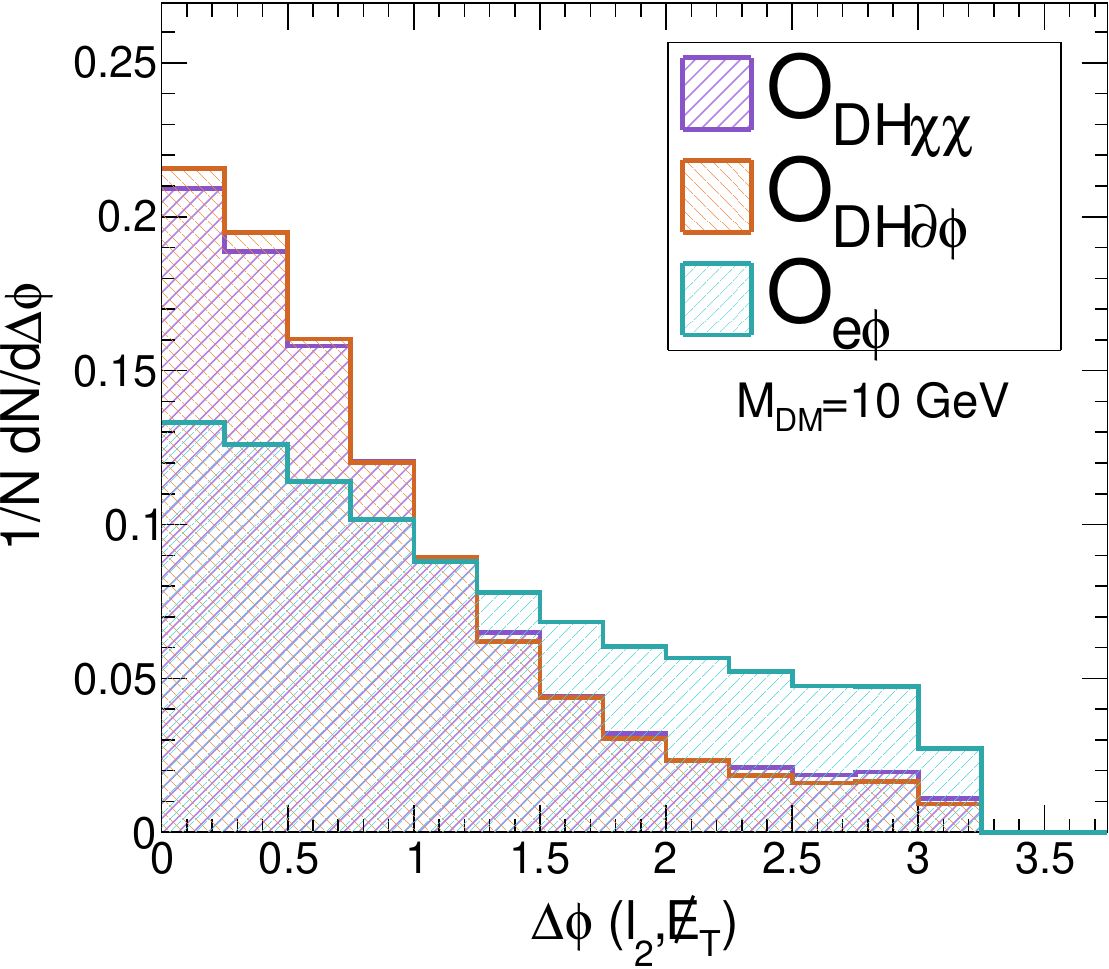}
        \caption{}
    \end{subfigure}

    \caption{\small 
    Distributions of the azimuthal angular separations 
    (a) $\Delta\phi(\ell_1,\ell_2)$, 
    (b) $\Delta\phi(\ell_1,\MET)$, and 
    (c) $\Delta\phi(\ell_2,\MET)$ 
    for selected operators. 
    All histograms are normalized to unit area and therefore show shape only. 
    We use $\wc=1$, $\Lambda=1~\textrm{TeV}$, and 
    $\sqrt{s}=14~\textrm{TeV}$ in all panels.}
    \label{fig:delta_phi}
\end{figure}

We consider two operator classes: (i) derivative currents
$H^\dagger \overleftrightarrow{D}^\mu H$ and (ii) Yukawa--type interactions
$\bar{l}_p e_r H$ coupling to the postulated new non-interacting particles.
These structures lead to distinct kinematic distributions
(Figures~\ref{fig:distributions_lep}--\ref{fig:distributions_lep2}).
We now examine whether LHC kinematics can discriminate between these two categories of operators.
As a benchmark, we take $M_{\rm DM} = 10~\text{GeV}$ and compare one
fermionic DM operator $\opdhchi$ and one scalar DM operator $\opdhphi$
with the Yukawa-type operator $\opephi$. In addition to the variables shown earlier, we study the azimuthal angles between the decay products of the Higgs boson. Fig.~\ref{fig:delta_phi} shows that $\Delta \phi(\ell_1, \ell_2)$ discriminates between  the derivative operators $\opdhchi$ and $\opdhphi$ and the Yukawa-like operators $\opephi$ very well. By contrast, the $\Delta \phi(\ell, \MET)$ distributions are much flatter, but still retain some discriminating power.
Together with Figures~\ref{fig:distributions_lep}-\ref{fig:distributions_lep2},
These basic azimuthal observables can help distinguish the two operator categories.

\begin{figure}[t]
    \centering
    \begin{subfigure}{0.47\textwidth}
        \centering
        \includegraphics[width=\linewidth]{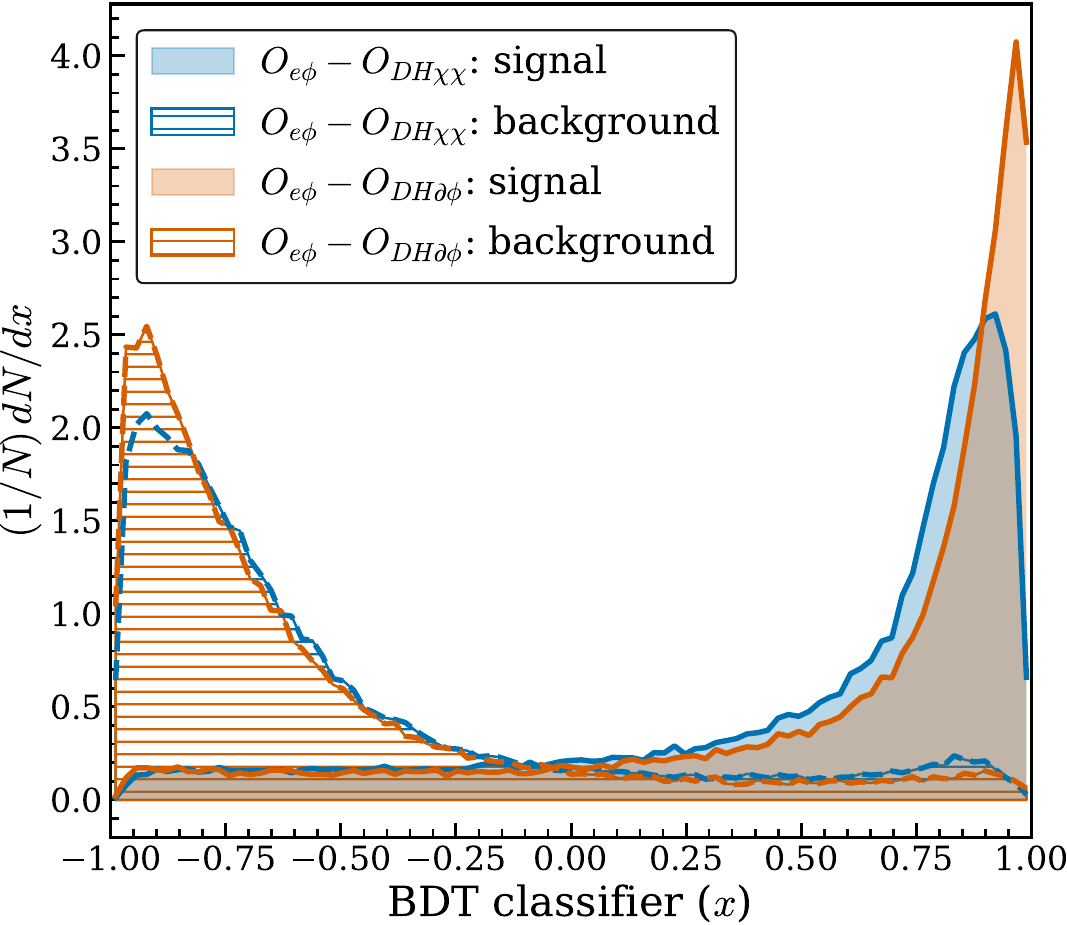}
        \caption{}
    \end{subfigure}
    \hfill
    \begin{subfigure}{0.47\textwidth}
        \centering \includegraphics[width=\linewidth]{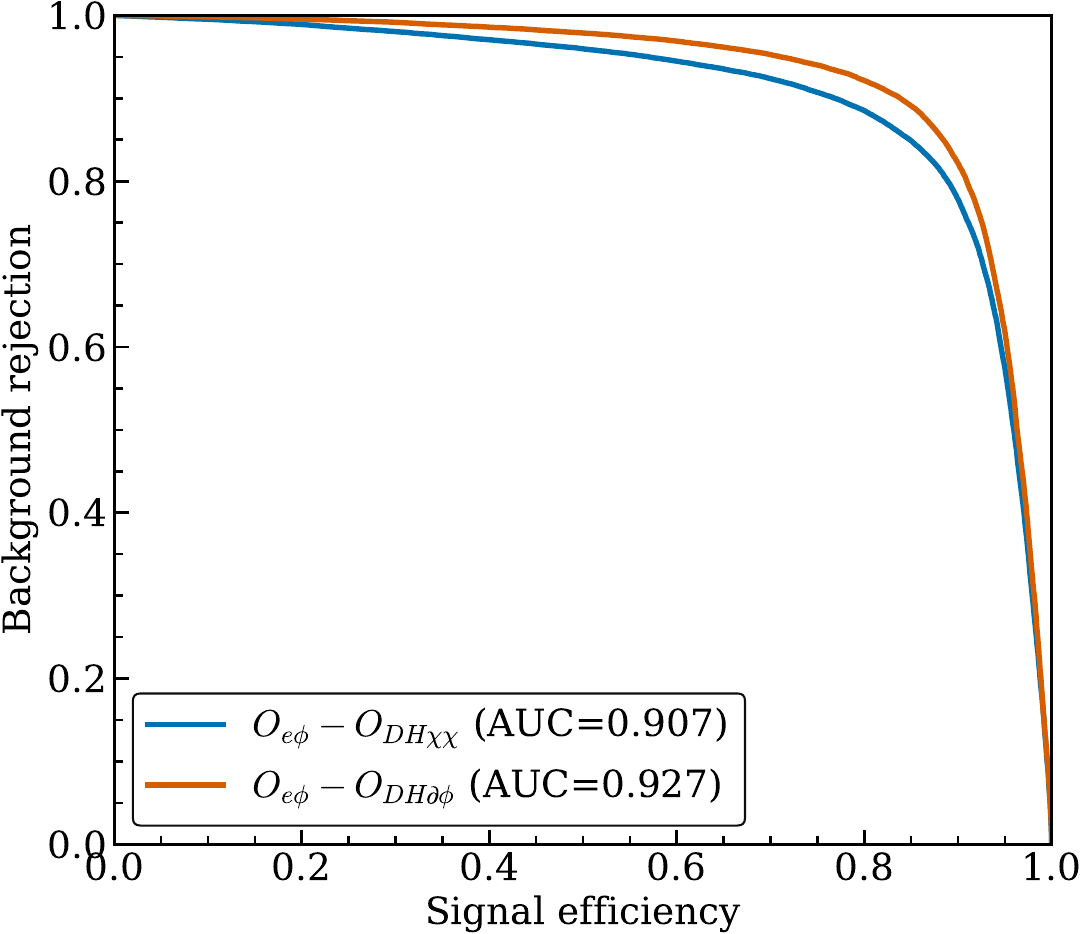}
        \caption{}
    \end{subfigure}
    \caption{\small 
    Distributions of 
    (a) the classifier response and 
    (b) the ROC curve, considering $\opephi$ as signal and 
    $\opdhphi$, $\opdhchi$ individually as backgrounds. 
    Events with $M_{\rm DM}=10~\textrm{GeV}$ are used for these plots, 
    with $\wc=1$, $\Lambda=1~\textrm{TeV}$, and 
    $\sqrt{s}=14~\textrm{TeV}$. 
    The AUC, Area Under the Curve, quantifies the performance of the BDT.}    \label{fig:tensor_op_discriminator}
\end{figure}

For a quantitative assessment, we train two separate MVAs, taking $\opephi$
as the ``signal'' and, in turn, $\opdhphi$ and $\opdhchi$ as the ``background''. The input kinematic variables include $M(\ell \ell)$, $M_T(\ell, \ell, \MET)$, $p_T(\ell_1)$, $p_T(\ell_2)$, $\MET$, and the three azimuthal angles $\Delta \phi(\ell_1, \ell_2)$, $\Delta \phi(\ell_1, \MET)$ and $\Delta \phi(\ell_2, \MET)$. As expected, the BDT identifies $M(\ell \ell)$, $M_T(\ell, \ell, \MET)$, and $\Delta \phi(\ell_1, \ell_2)$ as the most powerful discriminants, followed by the remaining variables.
The resulting classifier outputs and ROC curves are shown in
Fig.~\ref{fig:tensor_op_discriminator}. Overall, operators containing the current $H^\dagger \overleftrightarrow{D}^\mu H$
are efficiently separated from those with the Yukawa structure $\bar{l}_p e_r H$. This kind of kinematic discrimination among operator classes will be instrumental in inferring plausible UV completions if a new-physics signal is discovered.

\section{Results}
\label{sec:results}

\begin{figure}[t]
    \centering
    \begin{subfigure}{0.32\textwidth}
        \centering        \includegraphics[width=\linewidth]{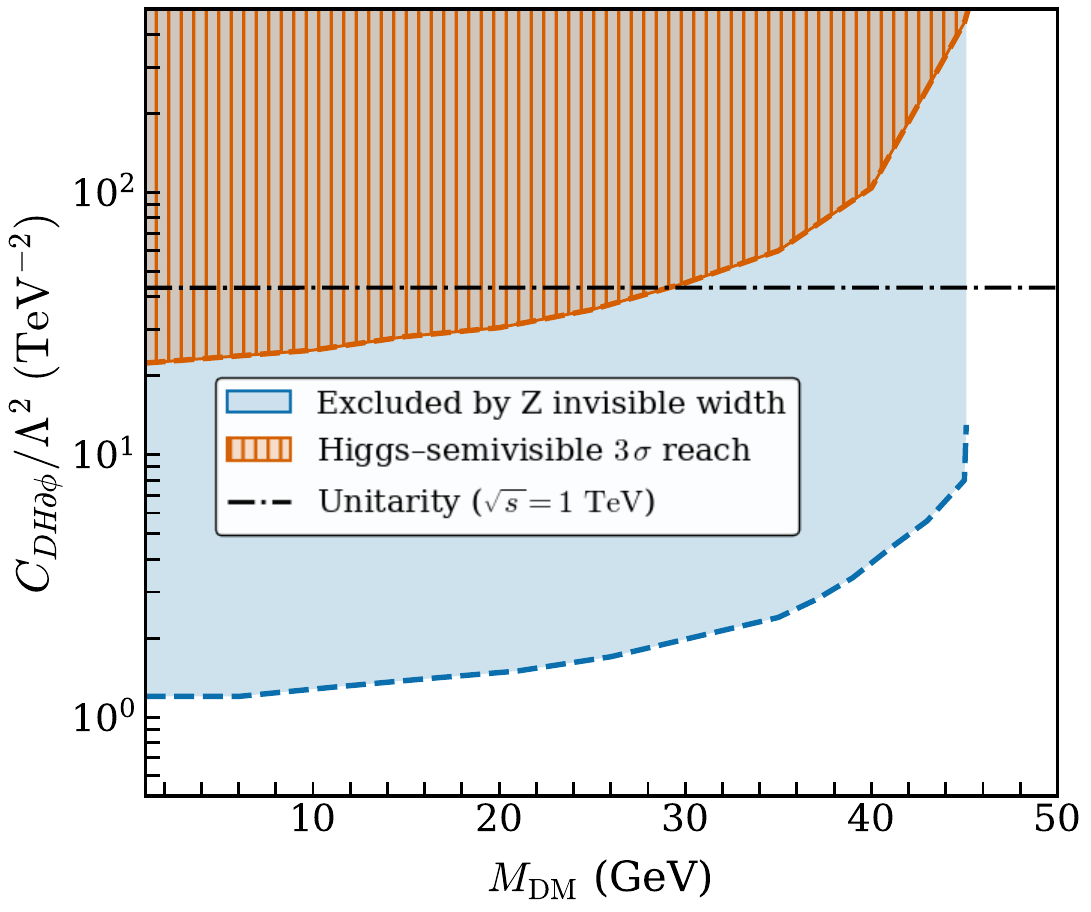}
        \caption{}
    \end{subfigure}
    \begin{subfigure}{0.32\textwidth}
        \centering        \includegraphics[width=\linewidth]{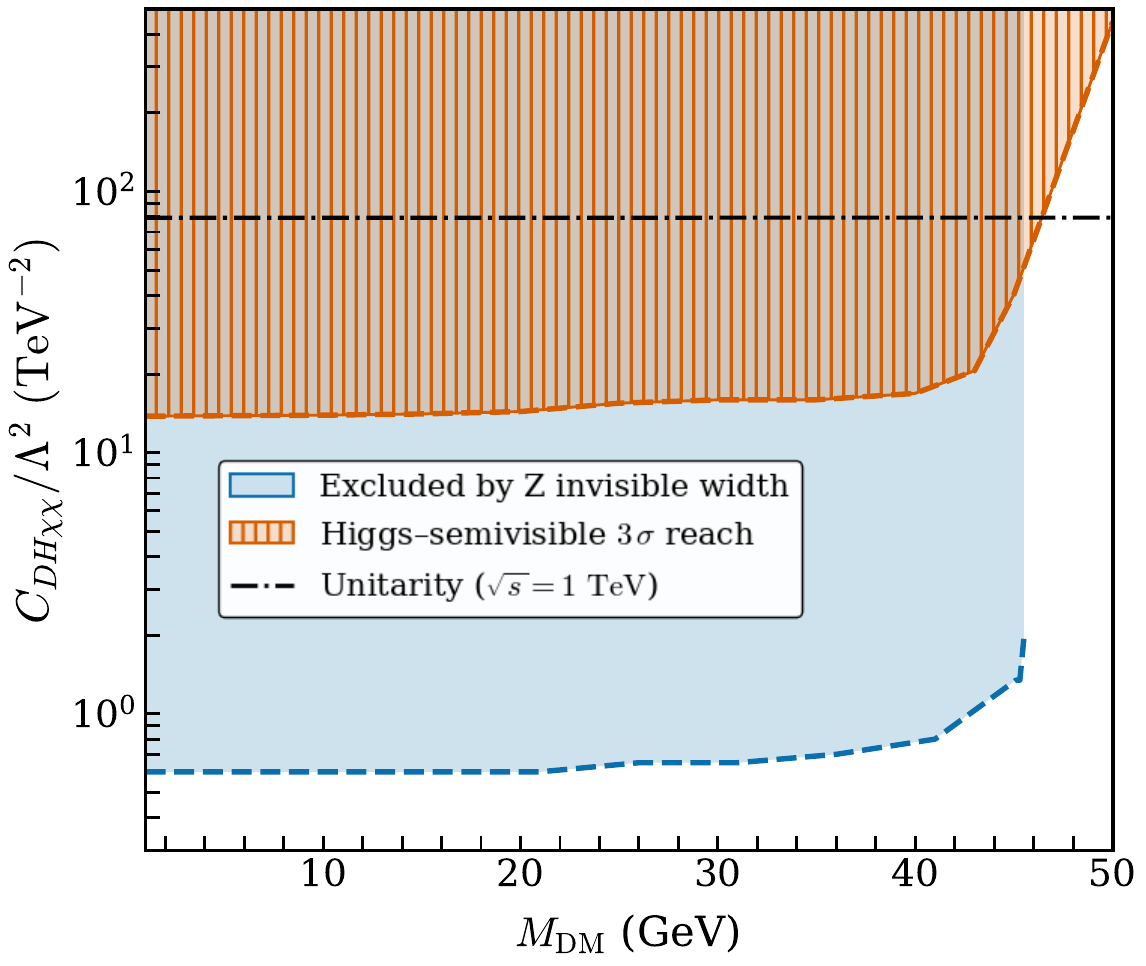}
        \caption{}
    \end{subfigure}
    \begin{subfigure}{0.32\textwidth}
        \centering        \includegraphics[width=\linewidth]{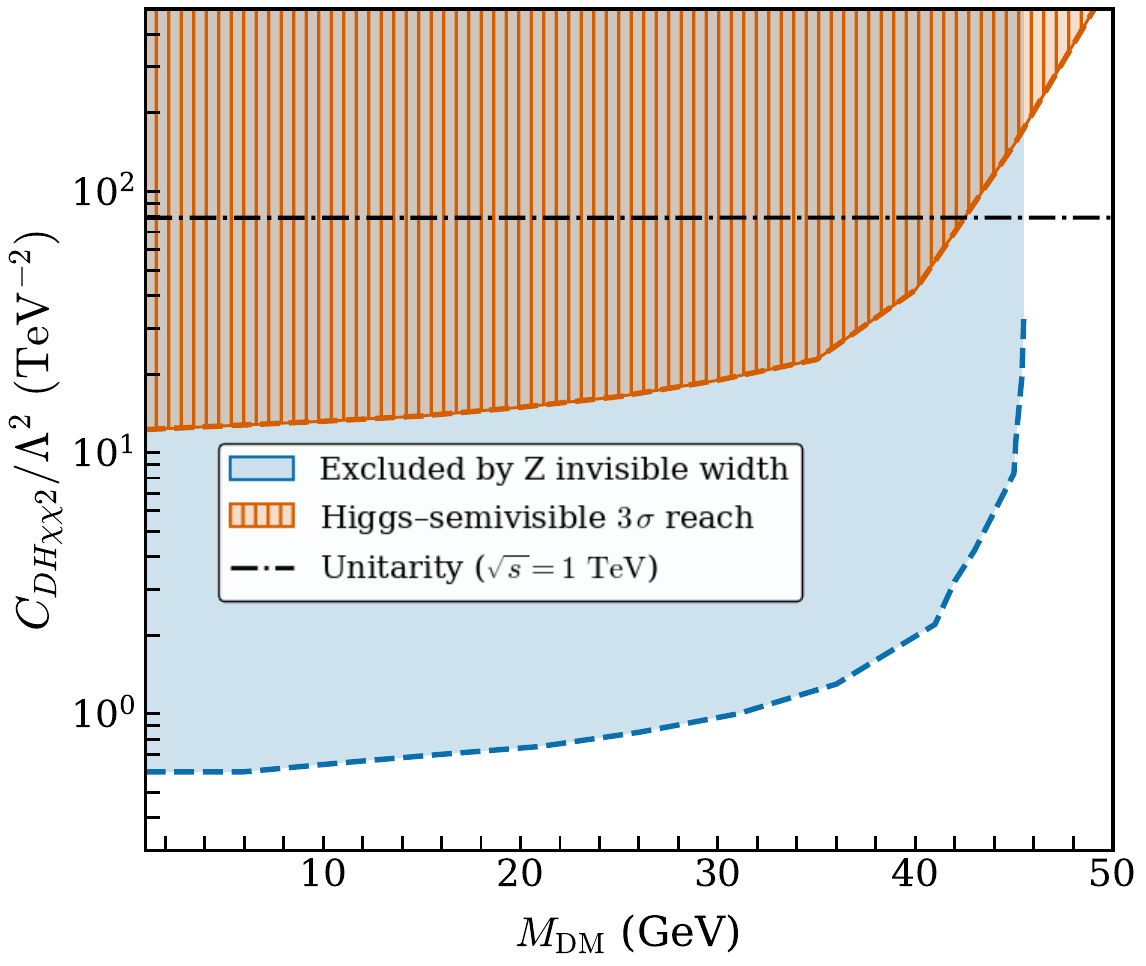}
        \caption{}
    \end{subfigure}
    \hspace{0.06\textwidth}
    \begin{subfigure}{0.32\textwidth}
        \centering        \includegraphics[width=\linewidth]{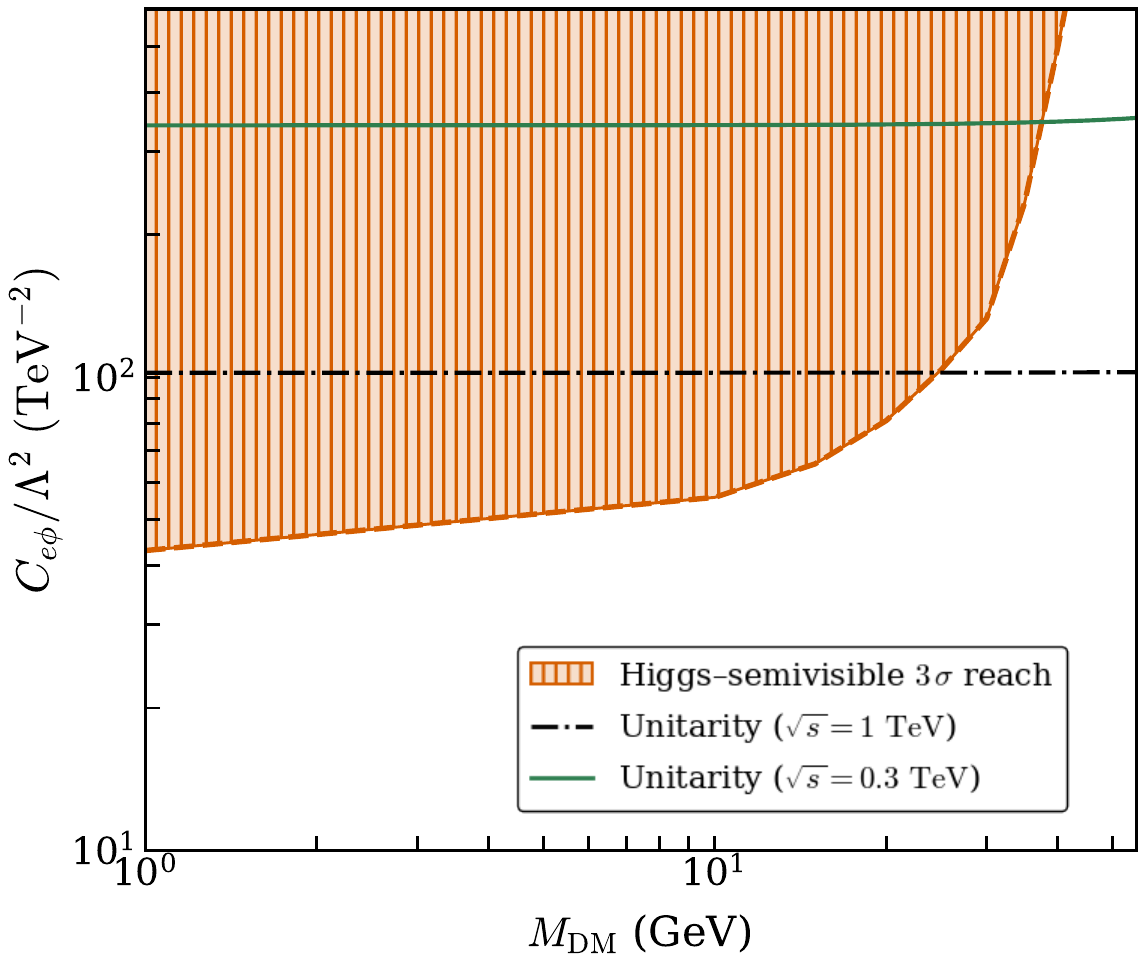}
        \caption{}
    \end{subfigure}
    \begin{subfigure}{0.32\textwidth}
        \centering        \includegraphics[width=\linewidth]{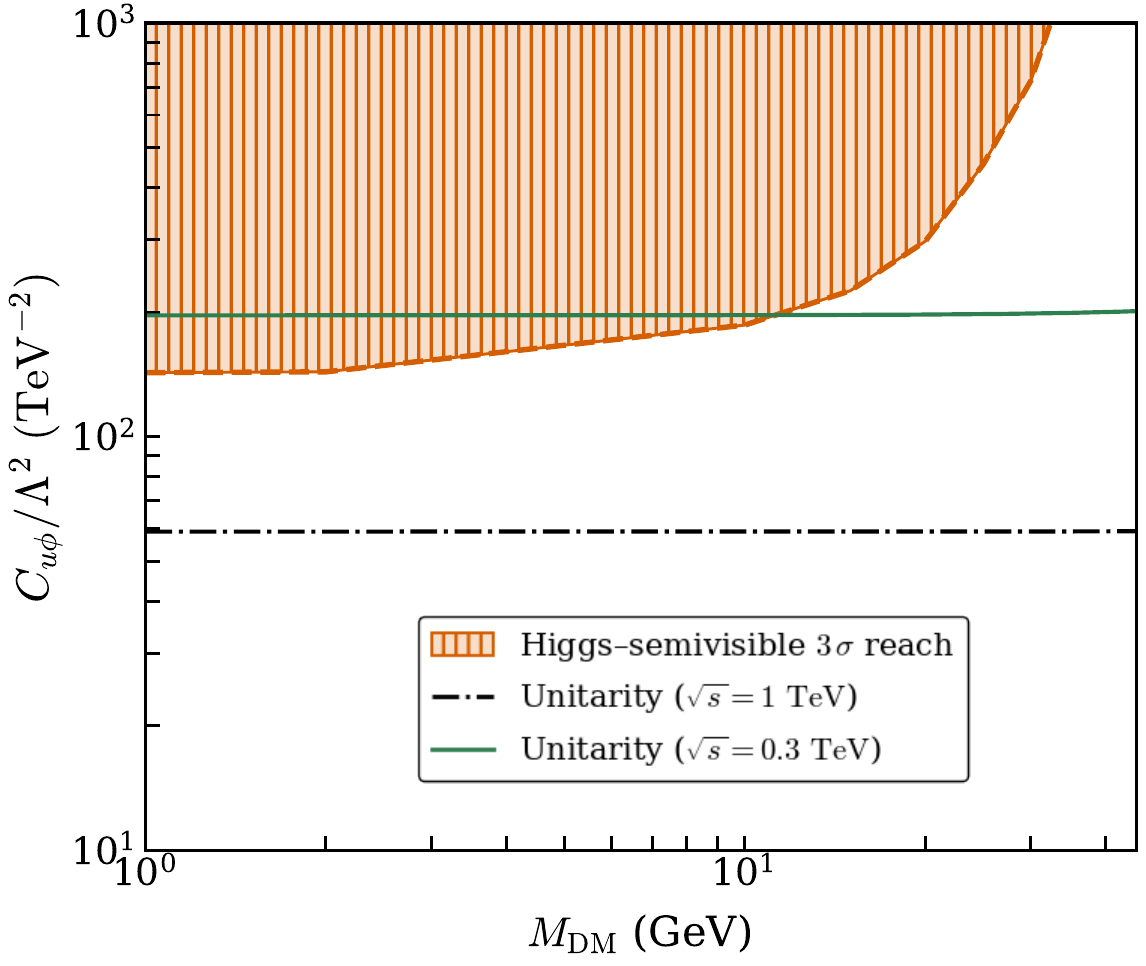}
        \caption{}
    \end{subfigure}
    \caption{\small 
    Regions excluded by the $Z$-boson invisible-decay limits at 95\% CL 
    (blue), obtained using Eq.~\ref{eq:z_inv_width}, and the $3\sigma$ 
    sensitivity to the various coefficients at the HL-LHC with 
    $\mathcal{L}=3000~{\rm fb}^{-1}$ and $\sqrt{s}=14~\textrm{TeV}$, 
    obtained from the semi-visible Higgs-decay multivariate analysis. 
    Here, we include only first- and second-generation fermions. 
    The unitarity bounds shown here correspond to 
    $\phi\phi\to\phi\phi$ scattering, Eqs.~\eqref{eq:unitscalar} 
    and \eqref{eq:unitfermion}, for panels (a)--(c), and to 
    $e^+e^-~(u\bar u)\to\phi\bar\phi$, Eq.~\eqref{boundoephi}, 
    for panels (d) and (e). 
    The sensitivity from semi-visible Higgs decays to $\opdphi$ 
    coincides with that of $\opuphi$.}
    \label{fig:bounds_leptonic}
\end{figure}

In Fig.~\ref{fig:bounds_leptonic} we present the HL-LHC $3\sigma$ sensitivity of the semi-visible Higgs decay analysis, with $\mathcal{L}=3000~\text{fb}^{-1}$ and $\sqrt{s}=14$ TeV alongside constraints from the $Z$-invisible width and partial-wave unitarity for the DSMEFT operators considered. For operators involving the derivative current $H^\dagger \overleftrightarrow{D}^\mu H$, the $Z$ invisible width provides the strongest bound up to $m_{\rm DM}\simeq45~\text{GeV}$, while the semi-visible Higgs analysis becomes competitive only for larger $m_{\rm DM}$. In contrast, for the operators $\opephi,\,\opuphi$, and $\opdphi$, the semi-visible Higgs decay analysis is the most sensitive. The partial-wave unitarity bounds are evaluated at $\sqrt{s}=1~\text{TeV}$ and $0.3~\text{TeV}$, the latter being close to the Higgs-decay scale. Most of the parameter space probed by the semi-visible Higgs analysis is consistent with unitarity at $\sqrt{s}=0.3~\text{TeV}$. However, at $\sqrt{s}=1~\text{TeV}$ the analysis remains sensitive only to a narrow region that satisfies unitarity for $\opdhphi,\,\opdhchi,\,\op{DH\chi\chi2}$ and $\opephi$, while it is fully excluded by unitarity for $\opuphi$ and $\opdphi$. The operators $\opuphi$ and $\opdphi$ are also constrained by monojet searches, where the bound found in  \cite{Roy:2025pht} translates to $\frac{{\cal C}_{u\phi,d\phi}}{\Lambda^{2}}\lesssim 3\rm \; TeV^{-2}$ for $M_{DM}\in[1,50]$ GeV. This is much stronger than the constraint that follows from this study, shown in Fig.~\ref{fig:bounds_leptonic}.  

\begin{figure}
    \centering    \includegraphics[width=0.65\linewidth]{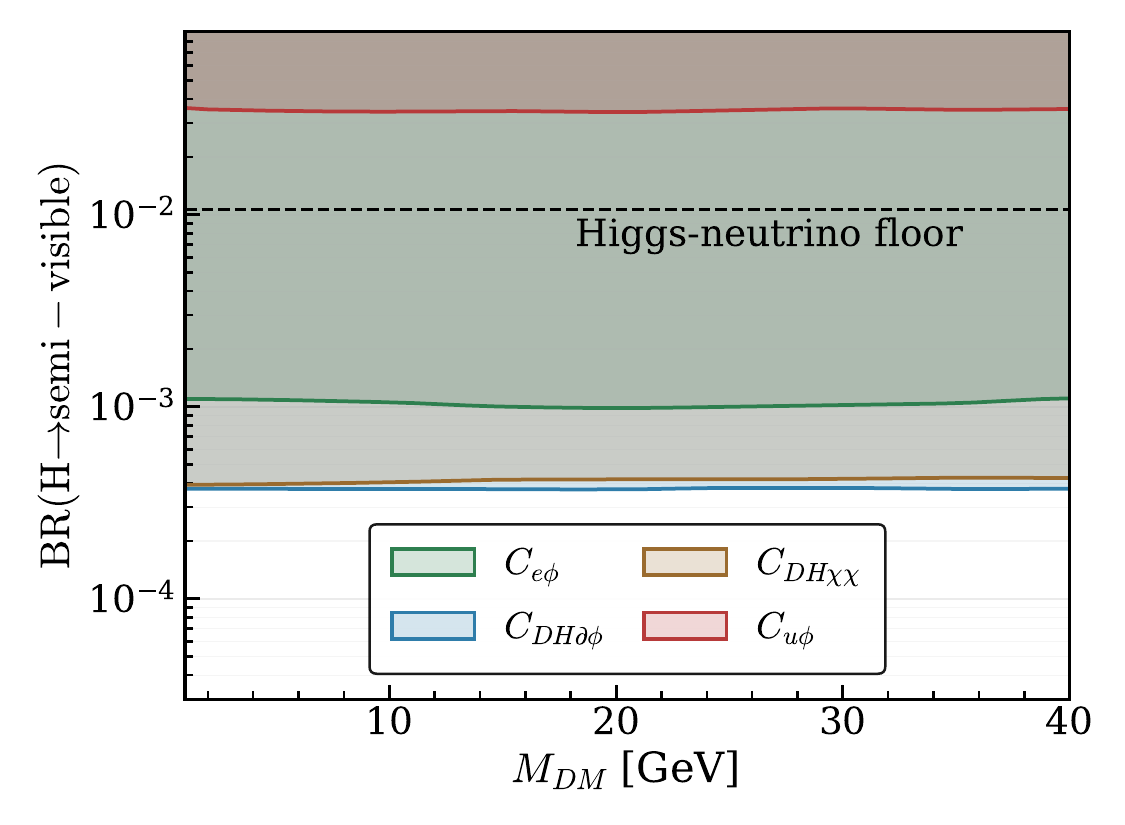}
    \caption{Level of the branching fraction $\rm BR(H\to semi\text{-}visible)$: $ BR(H\to \ell^+ \ell^- \, DM \, DM)$ or $BR(H\to jj \, DM \, DM)$ depending on the operator-type, where $DM=\phi$ or $\chi$,  that can be probed at 3$\sigma$ sensitivity for different values of $M_{DM}$ and different choices of operators at the HL-LHC with luminosity $\mathcal{L}=3000~\rm fb^{-1}$ and $\sqrt{s}=14$ TeV. The Higgs-neutrino floor curve (shown as the dashed black line) corresponds to  $BR(H \to \ell^+\ell^-\nu\bar{\nu})_{SM} = 1.06\times 10^{-2}$, and it would be about 30$\%$ lower for $BR(H \to qq\,\nu\bar{\nu})_{SM}$~\cite{LHCHiggsCrossSectionWorkingGroup:2016ypw}, which is relevant for $C_{u\phi}$. The sensitivities for $\op{DH\chi\chi2}$ and $\opdphi$ are the same as those for $\opdhchi$ and $\opuphi$, respectively, and are therefore not shown separately.
    }
    \label{fig:brsemivisible}
\end{figure}

In Fig.~\ref{fig:brsemivisible}, we present the region of sensitivity in terms of the semi-visible Higgs decay branching fraction as a function of the mass of the postulated DM particle (either $\phi$ or $\chi$).  We obtain the branching fraction by computing the partial widths $\Gamma(H\to \ell^+ \ell^- \, DM \, DM)$ or $\Gamma(H\to jj \, DM \, DM)$
(where $DM=\phi$ or $\chi$) for the regions shown in Fig.~\ref{fig:bounds_leptonic} using \texttt{MadGraph5}, and normalizing to the SM Higgs total width $\Gamma_{H}^{\rm tot}=4.1$ MeV \cite{LHCHiggsCrossSectionWorkingGroup:2016ypw}. The figure shows that for the operators that induce a semi-visible leptonic Higgs decay, it is possible to reach a sensitivity below the Higgs neutrino-floor, unlike with the models studied in \cite{Aguilar-Saavedra:2022xrb}, because, as shown above, this background is reducible here. Further comparing to the models in \cite{Aguilar-Saavedra:2022xrb}, we can reach sensitivities that are improved by one to two orders of magnitude due to the use of $M_{\ell\ell}$ as a discriminant. For the $\opuphi$ and $\opdphi$ operators, on the other hand, the sensitivity of a semi-visible Higgs decay analysis is comparatively much weaker. Although, nominally below the Higgs-neutrino floor, these bounds are observable as we have shown that this background is reducible in our case.

\section{Summary and Conclusion}
\label{sec:summary}

We have considered the set of dimension -6 effective operators that can give rise to semi-visible Higgs decays, through the production of pairs of new invisible scalars or fermions. Perturbative unitarity bounds on the coefficients of the effective operators, as well as constraints from the $Z$ invisible width, limit the size of the allowed interactions. We presented a framework for studying the semi-visible Higgs decays, utilizing simple kinematical cuts and contrasting with the results from a BDT. 

The main results are shown in Figures~\ref{fig:bounds_leptonic}~and~\ref{fig:brsemivisible}. The invisible $Z$ width provides the strongest constraints on $\wcdhphi~,\wcdhchi$ and $\wc_{DH\chi\chi2}$, whereas semi-visible Higgs decays are the most restrictive for $\wcephi$. Requiring perturbative unitarity to hold up to 1~TeV
provides the strongest bound on $\wcuphi$, indicating that the processes we study are not sensitive to the types of  BSM physics parameterized  by ${\cal O}_{u\phi}$ 
at scales above a few TeV. 
As  Figure \ref{fig:bounds_leptonic} also indicates, semi-visible Higgs decays constrain other scenarios where perturbative unitarity is satisfied for a narrow region with scale $\sim 1$ TeV, and almost everywhere for $\sqrt{s}\sim 0.3$~TeV, the scale probed by this decay. A BSM scenario described by these conditions is consistent with new degrees of freedom (which can be visible or invisible) appearing at around 0.5 TeV.
In Fig.~\ref{fig:brsemivisible}, we display the results in a different form, showing the Higgs branching fraction into the semi-visible channels that can be probed by HL-LHC. The figure also demonstrates that the region accessible to HL-LHC is below the Higgs-neutrino floor.

If the invisible final state is interpreted as thermal dark matter~\cite{Cowsik:1972gh,Lee:1977ua,Kolb:1990vq}, the semi-visible Higgs sensitivities are weaker than the limits from direct detection for hadronic operators \cite{LZ:2024zvo,PandaX:2022xqx,PandaX:2023xgl,DarkSide:2022dhx} and from indirect detection for leptonic operators \cite{ODonnell:2024aaw}. Nonetheless, they are complementary -- especially for different dark matter scenarios such as asymmetric dark matter, where indirect detection bounds are absent~\cite{Nussinov:1985xr,Gelmini:1986zz,Zurek:2013wia,Petraki:2013wwa} (see \cite{Roy:2024ear} for recent limits), or for co-annihilating dark matter, where both direct and indirect detection signals can be suppressed \cite{Bell:2013wua}. A quantitative study of semi-visible Higgs decay using DSMEFT for these scenarios 
will be the subject of future work.

\acknowledgments
\vspace{-5 mm}
This work is supported by an Australian Research Council Discovery Project. SD is supported by the U.S. Department of Energy under Contract No. DE- SC0012704. 
AR thanks Dibya S. Chattopadhyay, Biplob Bhattacherjee, and Saikat Karmakar for useful discussions, and Ranjan Laha for hosting him during his visit to IISc-CHEP, Bangalore, where part of this work was reviewed. Digital data is available from the authors upon request.

\bibliographystyle{utphys.bst}
\bibliography{ref.bib}
\end{document}

%% file: feynman.tex
\begin{figure}
	\centering
	\begin{subfigure}[b]{0.25\textwidth}
		\centering
		\begin{tikzpicture}
			\begin{feynman}
				\vertex [dot](c) at ( 1.9, 0.6) {};
				\vertex [dot](d) at ( 1.9, -0.6) {};
				\vertex (c1) at ( 2.7, 1.4) {$\bar{\nu}$};
				\vertex (c2) at ( 2.9, 0.8) {$\nu$};
				\vertex (d1) at ( 2.9, -0.8) {$\bar{f}$};
				\vertex (d2) at ( 2.7, -1.4) {$f$};
				\vertex[] (e) at (-0.75, 0){};
				\vertex[dot] (f) at ( 0.75, 0) {\contour{white}{}};
				\diagram* {
					(f) -- [photon,edge label=$\small{Z/Z^*}$] (c),
					(f) -- [photon, edge label=$\small{Z^{*}/Z}$] (d),
					(e) -- [scalar, edge label=$H$] (f),
					(c1) -- [fermion] (c),
					(c) -- [fermion] (c2),
					(d1) -- [fermion] (d),
					(d) -- [fermion] (d2)
				};
			\end{feynman}
		\end{tikzpicture}
        \caption*{~~~~~~~~(SM)}
	\end{subfigure}
    
    \begin{subfigure}[b]{0.25\textwidth}
		\begin{tikzpicture}
			\begin{feynman}
				\vertex [dot](d) at ( 1.9, -0.6) {};
				\vertex (c1) at ( 1.7, 1.2) {$\phi$};
				\vertex (c2) at ( 1.9, 0.6) {$\phi$};
				\vertex (d1) at ( 2.9, -0.8) {$\bar{f}$};
				\vertex (d2) at ( 2.7, -1.4) {$f$};
				\vertex[] (e) at (-0.75, 0){};
				\vertex[blob,scale=0.5] (f) at ( 0.75, 0) {\contour{white}{}};
				\diagram* {
					(f) -- [photon, edge label=$Z$] (d),
					(e) -- [scalar, edge label=$H$] (f),
					(f) -- [scalar] (c1),
					(f) -- [scalar] (c2),
					(d1) -- [fermion] (d),
					(d) -- [fermion] (d2)
				};
			\end{feynman}
		\end{tikzpicture}
        \caption*{~~~~~~~~~~~~(A)}
	\end{subfigure}
	\begin{subfigure}[b]{0.28\textwidth}
		\begin{tikzpicture}
			\begin{feynman}
				\vertex [blob,scale=0.5](c) at ( 1.9, 0.6) {};
				\vertex [dot](d) at ( 1.9, -0.6) {};
				\vertex (c1) at ( 2.7, 1.4) {$\phi$};
				\vertex (c2) at ( 2.9, 0.8) {$\phi$};
				\vertex (d1) at ( 2.9, -0.8) {$\bar{f}$};
				\vertex (d2) at ( 2.7, -1.4) {$f$};
				\vertex[] (e) at (-0.75, 0){};
				\vertex[dot] (f) at ( 0.75, 0) {\contour{white}{}};
				\diagram* {
					(f) -- [photon,edge label=$Z/Z^*$] (c),
					(f) -- [photon, edge label=$Z^{*}/Z$] (d),
					(e) -- [scalar, edge label=$H$] (f),
					(c) -- [scalar] (c1),
					(c) -- [scalar] (c2),
					(d1) -- [fermion] (d),
					(d) -- [fermion] (d2)
				};
			\end{feynman}
		\end{tikzpicture}
        \caption*{~~~~~~~~~~~~~~~~(B)}
	\end{subfigure}
		\begin{subfigure}[b]{0.28\textwidth}
		\centering
		\begin{tikzpicture}
			\begin{feynman}
				\vertex (c1) at ( 1.7, 1.2) {$\phi$};
				\vertex (c2) at ( 1.9, 0.6) {$\phi$};
				\vertex (d1) at ( 1.9, -0.6) {$\bar{f}$};
				\vertex (d2) at ( 1.7, -1.2) {$f$};
				\vertex[] (e) at (-0.75, 0){};
				\vertex[blob, scale=0.5] (f) at ( 0.75, 0) {\contour{white}{}};
				\diagram* {
					(d1) -- [fermion] (f),
					(f) -- [scalar] (c1),
					(f) -- [scalar] (c2),
					(f) -- [fermion] (d2),
					(e) -- [scalar, edge label=$H$] (f)
				};
			\end{feynman}
		\end{tikzpicture}
        \caption*{~~~~~~~~~~~~~~~~(C)}
	\end{subfigure}
	\caption{\small Representative Feynman diagrams illustrating semi-visible Higgs decays, independent of  the production mode. The top panel shows the SM process, while the bottom panel presents possible EFT-induced contributions. The cross-hatched circles represent insertions of DSMEFT operators. }
	\label{fig:feynmandiagrams}
\end{figure}